\documentclass[aps,prd,twocolumn,superscriptaddress,longbibliography,letterpaper]{revtex4}
\usepackage{amssymb}
\usepackage{cancel}
\usepackage{color,graphicx}
\usepackage{amsmath}
\usepackage{amsbsy}
\usepackage{amsthm}
\usepackage{bbm}
\usepackage{epsfig}
\usepackage{lscape}
\usepackage{float}
\usepackage{graphicx}
\usepackage{subfigure}
\usepackage{dcolumn}
\usepackage{bbm}
\usepackage{color,epstopdf}
\usepackage{amscd}
\usepackage{amsfonts}  
\usepackage[]{amsmath}
\usepackage{amssymb}    
\usepackage{mathrsfs}
\usepackage{verbatim}
\usepackage[]{cases}
\usepackage{amsmath}
\usepackage{wasysym}
\usepackage[utf8]{inputenc}
\usepackage[T1]{fontenc}
\usepackage{mathtools}
\usepackage[dvipsnames]{xcolor}
\usepackage{url}

\usepackage[colorlinks=true,linkcolor=blue,urlcolor=blue,citecolor=blue,pdfusetitle]{hyperref}
\usepackage{soul}

\usepackage{braket} %
\usepackage{bbm} %
\usepackage{colonequals} %
\usepackage[scr=boondoxo]{mathalfa} %

\usepackage{qcircuit} %
\usepackage{simplewick} %

\newcommand{\de}{\delta}

\newcommand{\tht}{\theta}  %

\newcommand{\la}{\lambda}
\newcommand{\rh}{\rho}
\newcommand{\si}{\sigma}
\newcommand{\ta}{\tau}

\newcommand{\ph}{\phi}
\newcommand{\ch}{\chi}

\newcommand{\om}{\omega}
\newcommand{\Ga}{\Gamma}

\newcommand{\La}{\Lambda}

\newcommand{\Ps}{\Psi}
\newcommand{\Om}{\Omega}
\newcommand{\cellcolour}{\cellcolour} %
\newcommand{\eqn}[1]{\begin{equation}#1\end{equation}} %
\newcommand{\aln}[1]{\begin{align}#1\end{align}} %
\newcommand{\subeqn}[1]{\begin{subequations}#1\end{subequations}} %

\newcommand{\pmx}[1]{\begin{pmatrix}#1\end{pmatrix}} %
\newcommand{\pw}[1]{\begin{cases}#1\end{cases}} %

\newcommand{\iu}{\mathrm{i}} %
\newcommand{\ec}{\mathrm{e}} %

\newcommand{\lb}{\left}  %
\newcommand{\rb}{\right}  %
\newcommand{\ct}{\dagger}  %
\newcommand{\infi}{\infty} %
\newcommand{\mc}{\mathcal} %
\newcommand{\Or}{\mathcal{O}} %
\newcommand{\nn}{\nonumber} %
\newcommand{\ti}{\widetilde} %
\newcommand{\ce}{\colonequals} %

\newcommand{\id}{\mathbbm{1}}
\newcommand{\tp}{\otimes} %
\newcommand{\bd}{\boldsymbol}

\newcommand{\intall}{\int_{-\infty}^{\infty}} %
\newcommand{\intallr}{\int_{0}^{\infty}} %
\newcommand{\qq}{\qquad\qquad} %
\newcommand{\qqqq}{\qquad\qquad\qquad\qquad} %
\newcommand{\tx}{\text} %

\newcommand{\sipl}{\hat{\sigma}^+} %
\newcommand{\simi}{\hat{\sigma}^-} %
\newcommand{\oad}{\hat{a}^{\dagger}} %
\newcommand{\Kb}[2]{\Ket{#1}\Bra{#2}} %
\newcommand{\KB}[3]{\Ket{#1}_{#2}\Bra{#3}} %
\newcommand{\Abs}[1]{\left|{#1}\right|} %

\DeclareMathOperator{\Tr}{Tr}

\DeclareMathOperator{\erf}{erf}
\DeclareMathOperator{\erfc}{erfc}
\DeclareMathOperator{\erfi}{erfi}

\DeclareMathOperator{\SI}{Si}
\DeclareMathOperator{\sinc}{sinc}

\makeatletter
\newcommand{\vast}{\bBigg@{3}}	
\newcommand{\Vast}{\bBigg@{4}}
\makeatother

\renewcommand{\Re}{\operatorname{Re}}
\renewcommand{\Im}{\operatorname{Im}}

\graphicspath{
  {"./NewFigures/"}
}

\let\originalleft\left
\let\originalright\right
\renewcommand{\left}{\mathopen{}\mathclose\bgroup\originalleft}
\renewcommand{\right}{\aftergroup\egroup\originalright}

\allowdisplaybreaks

\begin{document}

\title{Bandlimited Entanglement Harvesting}

\author{Laura J. Henderson}
\email[]{l7henderson@uwaterloo.ca}
\affiliation{Department of Physics and Astronomy, University of Waterloo, Waterloo, Ontario, Canada, N2L 3G1}
\affiliation{Institute for Quantum Computing, University of Waterloo, Waterloo, Ontario, Canada, N2L 3G1}
\affiliation{Centre for Quantum Computation and Communication Technology, School of Science, RMIT University, Melbourne, Victoria 3001, Australia}
\author{Nicolas C. Menicucci}
\email[]{ncmenicucci@gmail.com}
\affiliation{Centre for Quantum Computation and Communication Technology, School of Science, RMIT University, Melbourne, Victoria 3001, Australia}

\begin{abstract}
There are many reasons to believe that there is a fundamental minimum length scale below which distances cannot be reliably resolved.  One method of constructing a quantum field with a finite minimum length scale is to use bandlimited quantum field theory, where the spacetime is mathematically both continuous and discrete. This is a modification to the field, which has been shown to have many consequences at the level of the field. We consider an operational approach and use a pair of particle detectors (two-level qubits) as a local probe of the field, which are coupled to the vacuum of the bandlimited massless scalar field in a time dependent way through a switching function.  We show that mathematically, the bandlimit modifies the spatial profile of the detectors so that they are only quasi-local. We explore two different types of switching functions, Gaussian and Dirac delta. We find that with Gaussian switching, the bandlimit exponentially suppresses the de-excitation of the detectors when the energy gap between the two levels is larger than the bandlimit.  If the detectors are prepared in ground state, in certain regions of the parameter space they are able to extract more entanglement from the field than if there was no bandlimit.  When the detectors couple with Dirac-delta switching, we show that a particle detector is most sensitive to the bandlimit when it couples to a small but finite region of spacetime.  We find that the effects of a bandlimit are detectable using local probes. This work is important because it illustrates the possible observable consequences of a fundamental bandlimit in a quantum field. 
\end{abstract}
\date{December 9, 2020}
\maketitle

\section{Introduction}

One of the  most important  open problems of modern physics is the question of what is happening at the highest energy scales where quantum field theory and general relativity are expected to be incorporated into a larger theory of physics.  These high energy scales correspond to the shortest length scales, where it is widely thought that, due to quantum fluctuations of the metric, the notions of space and time break down \cite{Hossenfelder:2012lrr}.  Currently, there are many approaches to consistent theories of quantum gravity that describe the physics at these length scales, including string theory, loop quantum gravity and many others \cite{Rovelli:1998,Carlip:2015ijmpd}.

It has been shown that if there is a finite, minimum-length uncertainty in a quantum field, the field will obey the Shannon sampling theorem, meaning the continuous field can be reconstructed from a discrete set of sampling points \cite{Kempf:2000prl}.  This also implies the field will be bandlimited. The Shannon Sampling Theorem \cite{6773024} is a theory in classical information that provides an equivalence between a continuous and a discrete representation of information.  If a signal $f(t)$ is bandlimited, so that it has no frequencies higher than some $\Lambda$, then knowing the value of the signal $f(t_n)$ on a lattice of points $\{t_n\}$ is enough to reconstruct the continuous signal for all $t$ provided the average spacing of the lattice is less than or equal to the Nyquist spacing, $\pi/\Lambda$. Unlike naively putting a quantum field theory~(QFT) on a lattice, the resultant bandlimited QFT preserves local Euclidian symmetries, but is not Lorentz invariant.  However, the conventional momentum cutoff generalises to a covariant cutoff, where modes with a wavelength smaller than the cutoff have very small bandwidth and are effectively frozen out \cite{Kempf:2013jmp,Chatwin-Davies:2017prl}.

It was shown by Pye, Donnelly and Kempf \cite{PhysRevD.92.105022}, that applying a conventional bandlimit to a $(1+1)$-dimensional scalar QFT results in field degrees of freedom that occupy an incompressible spatial volume.  As a result, the two-point correlations of the field and entanglement entropy are modified from the case of no cutoff, with the least modification occurring at distances of the Nyquist spacing.

In that paper, the authors raised the idea of studying the interaction of Unruh-DeWitt (UDW) detectors \cite{PhysRevD.14.870}---two-level quantum systems---with bandlimited quantum fields as a natural next step.  The UDW detectors serve as  a local probe of the field.  This could be used as a stepping stone towards the quantisation of sampling theory.  More specifically, in the entanglement harvesting protocol, the spatial profile of the detector---which quantifies where the detector couples to the field---and the field degrees of freedom enter the model at the same level.  Exploring how they interact may provide a more operational understanding of the finite spatial volume of the discrete degrees of freedom.

In general, the study of UDW detectors as local particle detectors provides an operational method of probing quantum field fluctuations and correlations by providing a method of directly sampling the field in different spacetime regions through local interactions.  It has been shown by Valentini~\cite{VALENTINI1991321} and Reznik \emph{et al.}~\cite{Reznik:2002fz,PhysRevA.71.042104} that entanglement present in a quantum field can be extracted by a pair of initially separable UDW detectors that couple locally to the field in a process known as entanglement harvesting \cite{Salton:2014NJP}.

In most implementations of entanglement harvesting the setup is very simple: the detectors are generally coupled linearly to the vacuum of a scalar field.  However, it has been shown that this simple model is a good approximation to the light-matter interaction, under the assumption that no angular momentum is transferred \cite{PhysRevD.87.064038,PhysRevA.89.033835,PhysRevD.94.064074}.  Entanglement harvesting has gained interest in both its applicational and foundational potential \cite{Ralph:2015vra,Brown:2014njp,Salton:2014NJP,Martin-Martinez:2016,Rodriguez-Camargo:2016fbq,Richter:2017ljj,Richter:2017noq,Huang:2017yjt,Ardenghi:2018xrh}.  It may also have some implications for quantum gravity and the black hole information paradox \cite{Henderson:2017yuv,Pozas-Kerstjens:2017xjr,Ng:2018drz,Henderson:2018lcy,Ng:2018ilp,Sachs:2017exo,Simidzija:2018ddw,Trevison:2018cbf}.

In particular, this protocol has been shown to be incredibly sensitive to the properties of field to which they are coupled.  In particular, it has been shown to be able to distinguish between a thermal bath and a de~Sitter spacetime at the same temperature \cite{PhysRevD.79.044027,MartinMartinez:2012sg,Martin-Martinez:2014gra}, and it is sensitive to the topology of spacetime \cite{Smith:2016prd,Ng:2017prd}, the presence of horizons \cite{Henderson:2017yuv} and the boundary conditions \cite{CongTjoa:2019jhep} of the field. UDW detectors have also been shown to be very sensitive to their state of motion \cite{Salton:2014NJP,Brown:2014njp}.  %
There are generally two different assumptions made about the interaction between the detectors and the field.  First, and more commonly, it is assumed that the detectors couple only weakly with the field.  This allows for use of perturbation theory when calculating the final state of the system.  In this regime, entanglement harvesting is possible, but multi-phonon interactions are not considered.  Additionally, this model may lead to divergences in limits where the detector response becomes large but finite.  The second assumes that the detectors have Dirac-delta switching---i.e.,\ they couple to the field at single moment in time.  Here, the time evolution operator takes a much simpler form, and in some cases the final state of the system can be known exactly.  However, entanglement harvesting is not possible with delta switching \cite{Simidzija:2017kty,PhysRevD.97.125002}.

In this paper we implement a conventional bandlimit on the scalar QFT by applying a hard cutoff and only allow modes with frequencies $|\bd{k}|<\Lambda$ to propagate.  Such a cutoff is not Lorentz invariant, but we expect that a similar result will hold in the case of a covariant cutoff, since our UDW detectors only couple to the field for a finite time, and modes with a small bandwidth would be further suppressed.

The rest of the paper is as follows. In section \ref{sec:UDW}, we review the UDW model and use perturbation theory to find the final state of the two detectors to lowest order in the coupling strength.  Before specifying the dimensionality, the spatial profile or switching function, we introduce the notion of the effective spatial profile, where we have absorbed the momentum cutoff of the field into a nonlocal modified spatial profile, allowing for a non-bandlimited field to model the same physics (if the modified profile is used instead of the original).  Next, in section \ref{sec:Setup}, we specify our model and take the detectors to be point like and couple to the field with a Gaussian function switching. We perturbatively calculate the excitation and de-excitation probability of a single detector in section \ref{sec:TP} and the entanglement harvested by a pair of detectors in section \ref{sec:EH}. We use Dirac-delta switching to non-perturbatively calculate the transition probability of two detectors with Gaussian spatial profiles in section \ref{sec:delta}. Finally, in section \ref{sec:Conclusion}, we conclude and discuss potential future work.
\section{The Unruh-DeWitt model}
\label{sec:UDW}

We will use UDW detectors to model the two particle detectors, $A$ and $B$, which capture most of the features of the light-matter interaction when no angular momentum is exchanged. In this model, the detectors are described by two-level quantum systems, with ground and excited states given by $\Ket{0}_D$ and $\Ket{1}_D$, respectively, and separated by an energy gap of $\Om_D$, which couple locally to a quantum scalar field~$\hat{\phi}(\bd{x},t)$, with ${D \in \{A,B\}}$.  The interaction of each detector is described, in the interaction picture, by the Hamiltonian
\aln{
  \hat{H}_D(\ta_D) &= \la \ch_D(\ta_D)\lb(\ec^{\iu\Om_D\ta_D}\sipl_D + \ec^{-\iu\Om_D\ta_D}\simi_D\rb) \nn\\
  &\qquad \tp \int d^n\bd{x}\ F_D[\bd{x}-\bd{x}_D(\ta_D)]\hat{\ph}(\bd{x},t)
  \label{eq:HIntD}
}
where $\la$ is the coupling strength of the interaction, $\ch_D(\ta_D)$ is the switching function, which controls the interaction time, $\sipl_D \ce \KB{1}{D}{0}$ and $\simi_D \ce \KB{0}{D}{1}$ are the $\tx{SU}(2)$ ladder operators acting on the Hilbert space of detector $D$, $F_D(\bd{x})$ describes the spatial profile of the detector, and $\bd{x}_D(\ta_D)$ is the detector's spacetime trajectory parameterised by its proper time, $\ta_D$.

The time evolution of the detector-field system with respect to the time $t$ is generated by the unitary operator
\eqn{
  \hat{U} \ce \mc{T}\exp\lb[-\iu\int dt\ \lb(\frac{d\ta_A}{dt}\hat{H}_A[\ta_A(t)]+\frac{d\ta_B}{dt}\hat{H}_B[\ta_B(t)]\rb)\rb]
  \label{eq:TimeEvOp}
}
where $\mathcal{T}$ is the time ordering operator.

Following the entanglement harvesting protocol, we consider the detectors to be initially $(t\to-\infty)$ in the ground state and the field initially in the vacuum, so the joint state of the system is
\eqn{
  \Ket{\Ps_0} = \Ket{0}_A\tp\Ket{0}_B\tp\Ket{0}_\ph
}
and after the interaction $(t\to\infty)$, the system is in the state
\eqn{
  \Ket{\Ps_f} = \hat{U}\Ket{\Ps_0} = \sum_n \hat{U}^{(n)}\Ket{\Ps_0}
}
where $\hat{U}^{(n)}$ is the $n^\tx{th}$ term in the Dyson expansion of the time evolution operator (\ref{eq:TimeEvOp})
\eqn{
  \hat{U}^{(n)} = (-\iu)^n \int_{t_n<\dotsb<t_1} dt_1\dotsb dt_n\ \hat{H}(t_1)\dotsb\hat{H}(t_n)
  \label{eq:USeries}
}
and $\hat{H}(t) = \sum_{D\in\{A,B\}}\frac{d\ta_D}{dt}\hat{H}_D[\ta_D(t)]$.

We are only interested in the partial state of the two detectors, so the Hilbert space of the field is traced out:
\aln{
  \hat{\rh}_{AB} &\ce \Tr_\ph\big[\Kb{\Ps_f}{\Ps_f}\big] \nn\\
  &= \sum_{m,n}\Tr_\ph\big[\hat{U}^{(m)}\Kb{\Ps_0}{\Ps_0}(\hat{U}^{(n)})^\ct\big].
  \label{eq:rhoABSeries}
}
It can be seen from equations \ref{eq:HIntD} and \ref{eq:USeries} that in the expression $\hat{U}^{(n)}\Ket{\Ps_0}$, the field operator~$\hat{\ph}(\bd{x},t)$ will be applied to the vacuum $n$ times, and so the only terms that survive the partial trace and contribute to the partial state are terms where $m$ and $n$ have the same parity.  Additionally, in the final state, $\hat{U}^{(n)}\Ket{\Ps_0}$, an even $n$ results in both detectors being excited or both remaining in the ground state, and an odd $n$ results in only one detector being excited. Since $m$ and $n$ must be both even or both odd, the reduced density matrix of the two detectors will be of the form
\aln{
  \hat{\rh}_{AB} = \pmx{
    \rh_{11} & 0 & 0 & \rh_{14}\\
    0 & \rh_{22} & \rh_{23} & 0\\
    0 & \rh_{23}^* & \rh_{33} & 0\\
    \rh_{14}^* & 0 & 0 & \rh_{44}
  }
  \label{eq:rhoABGeneral}
}
in the basis \{${\Ket{0}_A\tp\Ket{0}_B}$, ${\Ket{0}_A\tp\Ket{1}_B}$, ${\Ket{1}_A\tp\Ket{0}_B}$, ${\Ket{1}_A\tp\Ket{1}_B}$\}.

The bandlimit~$\Lambda$ is a hard cutoff and is implemented by expanding the field operator in plane-wave modes and cutting off momenta where $\Abs{\bd{k}}>\La$:
\eqn{
  \hat{\ph}_\La(\bd{x},t) = \frac{1}{(2\pi)^{n/2}} \int_{\Abs{\bd{k}}<\La} \frac{d^n\bd{k}}{\sqrt{2\omega_{\bd{k}}}} \lb(\ec^{\iu(\omega_{\bd{k}}t-\bd{k}\cdot\bd{x})} \oad_{\bd{k}} + \text{H.c.} \rb),
  \label{eq:PWwithCutoff}
}
where, for a massless field the dispersion relation is $\omega_{\bd{k}}=\Abs{\bd{k}}$.
This implementation of the cutoff is not covariant but is chosen this way for ease of calculation and to follow on from the framework of~\cite{PhysRevD.92.105022}. Reproducing these calculations with a fully covariant cutoff \cite{Kempf:2013jmp} is left to future work. With this field operator, the interaction Hamiltonian of a single detector becomes
\aln{
  &\hat{H}_D(\ta_D) = \la\ch(\ta_D)\lb(\ec^{\iu\Om_D\ta_D}\sipl_D+\ec^{-\iu\Om_D\ta_D}\simi_D\rb) \nn\\
  &\quad \tp\int_{\Abs{\bd{k}}<\La} \frac{d^n\bd{k}}{\sqrt{2\Abs{\bd{k}}}} \lb(\ti{F}_D^*(\bd{k}) \ec^{\iu[\Abs{\bd{k}}t(\ta_D)-\bd{k}\cdot\bd{x}_D(\ta_D)]}\oad_{\bd{k}} + \tx{H.c.}\rb)
  \label{eq:HwithCutoff}
}
where
\eqn{
  \ti{F}_D(\bd{k}) = \frac{1}{(2\pi)^{n/2}} \int d^n\bd{x}\ F_D(\bd{x})\ec^{\iu\bd{k}\cdot\bd{x}}
}
is the Fourier transform of the spatial profile.

Mathematically, as was shown in \cite{grimmer2019machine}, the hard momentum cutoff in the interaction Hamiltonian (\ref{eq:HwithCutoff}) can be absorbed into the definition of the spatial profile:
\begin{widetext}
\aln{
  \hat{H}_D(\ta_D) &= \la\ch(\ta_D)\lb(\ec^{\iu\Om_D\ta_D}\sipl_D+\ec^{-\iu\Om_D\ta_D}\simi_D\rb)\tp\int \frac{d^n\bd{k}}{\sqrt{2\Abs{\bd{k}}}} \Pi\lb(\frac{\Abs{\bd{k}}}{2\La}\rb) \lb(\ti{F}_D^*(\bd{k}) \ec^{\iu[\Abs{\bd{k}}t(\ta_D)-\bd{k}\cdot\bd{x}_D(\ta_D)]}\oad_{\bd{k}} + \tx{H.c.}\rb) \nn\\
  &= \la\ch(\ta_D)\lb(\ec^{\iu\Om_D\ta_D}\sipl_D+\ec^{-\iu\Om_D\ta_D}\simi_D\rb) \tp \int \frac{d^n\bd{k}}{\sqrt{2\Abs{\bd{k}}}}\lb(\ti{G}_D^*(\bd{k})\ec^{\iu[\Abs{\bd{k}}t(\ta_D)-\bd{k}\cdot\bd{x}(\ta_D)]}\oad_{\bd{k}}+\tx{H.c.}\rb)
  \label{eq:HwithG}
}
\end{widetext}
where the new spatial profile is
\aln{
  G(\bd{x}) = F(\bd{x}) * \frac{1}{(2\pi)^{n/2}} \int d^n\bd{k}\ \Pi\lb(\frac{\Abs{\bd{k}}}{2\La}\rb) \ec^{-\iu\bd{k}\cdot\bd{x}}
  \label{eq:NewProfile}
}
and $\Pi(k)$ is the rectangle function
\eqn{
  \Pi(k) \ce \pw{
    0, & |k| > \frac{1}{2} \\
    1/2, & |k| = \frac{1}{2} \\
    1, & |k| < \frac{1}{2}
  }
}
In 3 dimensions, its inverse Fourier transform is
\eqn{
  \mc{F}^{-1}\lb[\Pi\lb(\frac{\Abs{\bd{k}}}{2\La}\rb)\rb]
  = \sqrt{\frac{2}{\pi}}\La^2\frac{j_1\big(\La\Abs{\bd{x}}\big)}{\Abs{\bd{x}}}
  \label{eq:thinc}
}
where $j_1(r)$ is the order-1 spherical Bessel function of the first kind. 

By writing the interaction Hamiltonian in this way, it is made clear that an alternative interpretation of the hard bandlimit $\Abs{\bd{k}}<\La$ is the pair of detectors that interact with a non-bandlimited $(\La\to\infi)$ massless scalar field have the spatial profile given by equation \ref{eq:thinc}.  Since the rectangle function is compact in $k$-space, the modified spatial profile is not compact---even if the initial profile~$F(\bd{x})$ was compact or even point-like.

Substituting the interaction Hamiltonian given by equation \ref{eq:HwithCutoff} into the equation \ref{eq:rhoABSeries} and expanding to the lowest order in the interaction strength gives the reduced density matrix of the two detectors
\begin{widetext}			
\eqn{
  \hat{\rh}_{AB} =  \pmx{
    1-P_A-P_B & 0 & 0 & X^* \\
    0 & P_B & C^* & 0 \\
    0 & C & P_A & 0\\
    X & 0 & 0 & 0
  } + \Or(\la^4)
  \label{eq:rhABPerturb}
}
where
\aln{
  P_D &\ce \frac{\la^2}{(2\pi)^n} \int \frac{d^n\bd{k}}{2\Abs{\bd{k}}} \Abs{\intall d\ta_D\ \ti{F}_D^*(\bd{k})\ch_D(\ta_D) \ec^{\iu[\Om_D\ta_D+\Abs{\bd{k}}t(\ta_D)-\bd{k}\cdot\bd{x}_D(\ta_D)]}}^2 \label{eq:PDGeneral}\\
  C&\ce \frac{\la^2}{(2\pi)^n} \int \frac{d^n\bd{k}}{2\Abs{\bd{k}}} \bigg[\lb(\intall d\ta_A\ \ti{F}_A^*(\bd{k}) \ch_A(\ta_A) \ec^{\iu[\Om_A\ta_A+\Abs{\bd{k}}t(\ta_A)-\bd{k}\cdot\bd{x}_A(\ta_A)]}\rb) \nn\\
  &\qqqq \times \lb(\intall d\ta_B\ \ti{F}_B^*(\bd{k}) \ch_B(\ta_B) \ec^{\iu[\Om_B\ta_B+\Abs{\bd{k}}t(\ta_B)-\bd{k}\cdot\bd{x}_B(\ta_B)]}\rb)^*\bigg] \\
  X &\ce \frac{\la^2}{(2\pi)^n} \int \frac{d^n\bd{k}}{2\Abs{\bd{k}}} \intall dt \int_{-\infi}^{t}dt'\ \ec^{-\iu\Abs{\bd{k}}(t-t')}  \nn\\
  &\qq \times \bigg[\bigg( \frac{d\ta_A}{dt}\frac{d\ta_B}{dt'} \ti{F}_A(\bd{k})\ti{F}_B^*(\bd{k}) \ch_A\big(\ta_A(t)\big)\ch_B\big(\ta_B(t')\big)\ec^{\iu[\Om_A\ta_A(t)+\Om_B\ta_B(t')]} \ec^{\iu\bd{k}\cdot[\bd{x}_A(t)-\bd{x}_B(t')]}\bigg) \nn\\
  &\qqqq + \bigg(\frac{d\ta_B}{dt} \frac{d\ta_A}{dt'} \ti{F}_B(\bd{k})\ti{F}_A^*(\bd{k}) \ch_B\big(\ta_B(t)\big) \ch_A\big(\ta_A(t')\big) \ec^{\iu[\Om_B\ta_B(t)+\Om_A\ta_A(t')]} \ec^{\iu\bd{k}\cdot[\bd{x}_B(t)-\bd{x}_A(t')]}\bigg)\bigg] \label{eq:XGeneral} %
}
\end{widetext}
and $D,D'\in\{A,B\}$.  Tracing out either detector $B$ or detector $A$ gives the reduced density matrix for the remaining detector
\aln{
  &\hat{\rh}_{A} = \Tr_B\lb[\hat{\rh}_{AB}\rb] = \pmx{
    1 - P_A & 0 \\
    0 & P_A
  } \nn\\
  &\hat{\rh}_{B} = \Tr_A\lb[\hat{\rh}_{AB}\rb] = \pmx{
    1 - P_B & 0 \\
    0 & P_B
  }
}
respectively, so we can interpret $P_D$ as the transition probability of detector $D$.  To leading order, the term $X$ encodes the non-local correlations between the two detectors, and the term $C$ encodes the total correlations \cite{Sachs:2017exo}.  

In addition to quantifying effect of the bandlimit on the transition probability of the detectors, we also wish to quantify its effect on the amount of entanglement the pair extracts from the vacuum.  We will use negativity \cite{Vidal:2002pra}~as our measure of entanglement.  The negativity is an entanglement monotone and is defined as
\eqn{
  \mc{N}\lb(\hat{\rh}_{AB}\rb) \ce \frac{\lb\|\hat{\rh}_{AB}^{\Ga_A}\rb\|-1}{2} = \sum_{\la_i<0}\Abs{\la_i}
  \label{eq:neg}
}
where $\Ga_A$ denotes the partial transpose with respect to $A $, $\|\cdot\|$ is the trace norm and the sum is over the negative eigenvalues of $\hat{\rh}_{AB}^{\Ga_A}$. For a two-qubit system like ours (and also for a qubit-qutrit system), the negativity is zero if and only if the state is separable.

\begin{figure*}[t]
  \centering
  \includegraphics[width=0.49\linewidth]{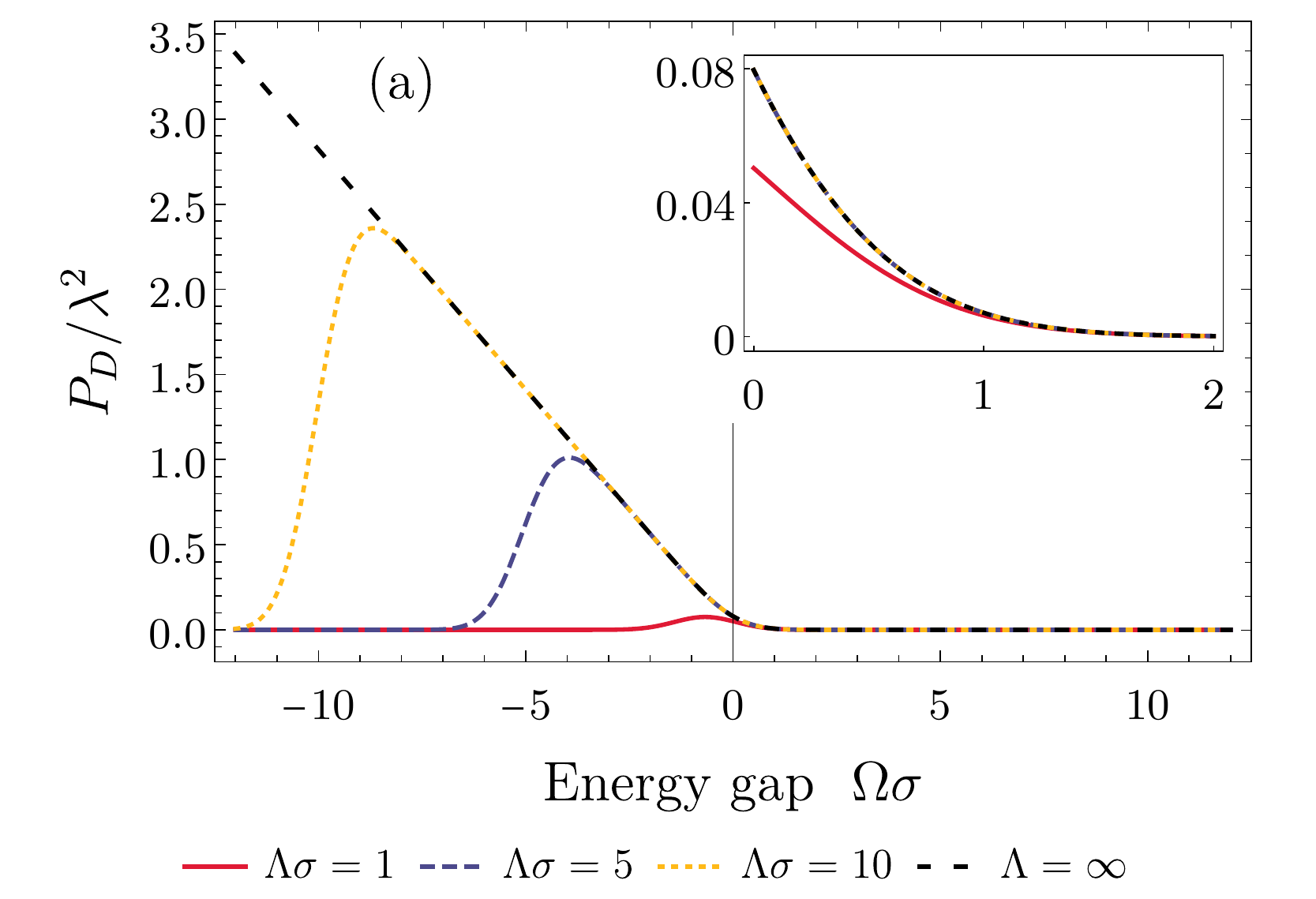}
  \includegraphics[width=0.49\linewidth]{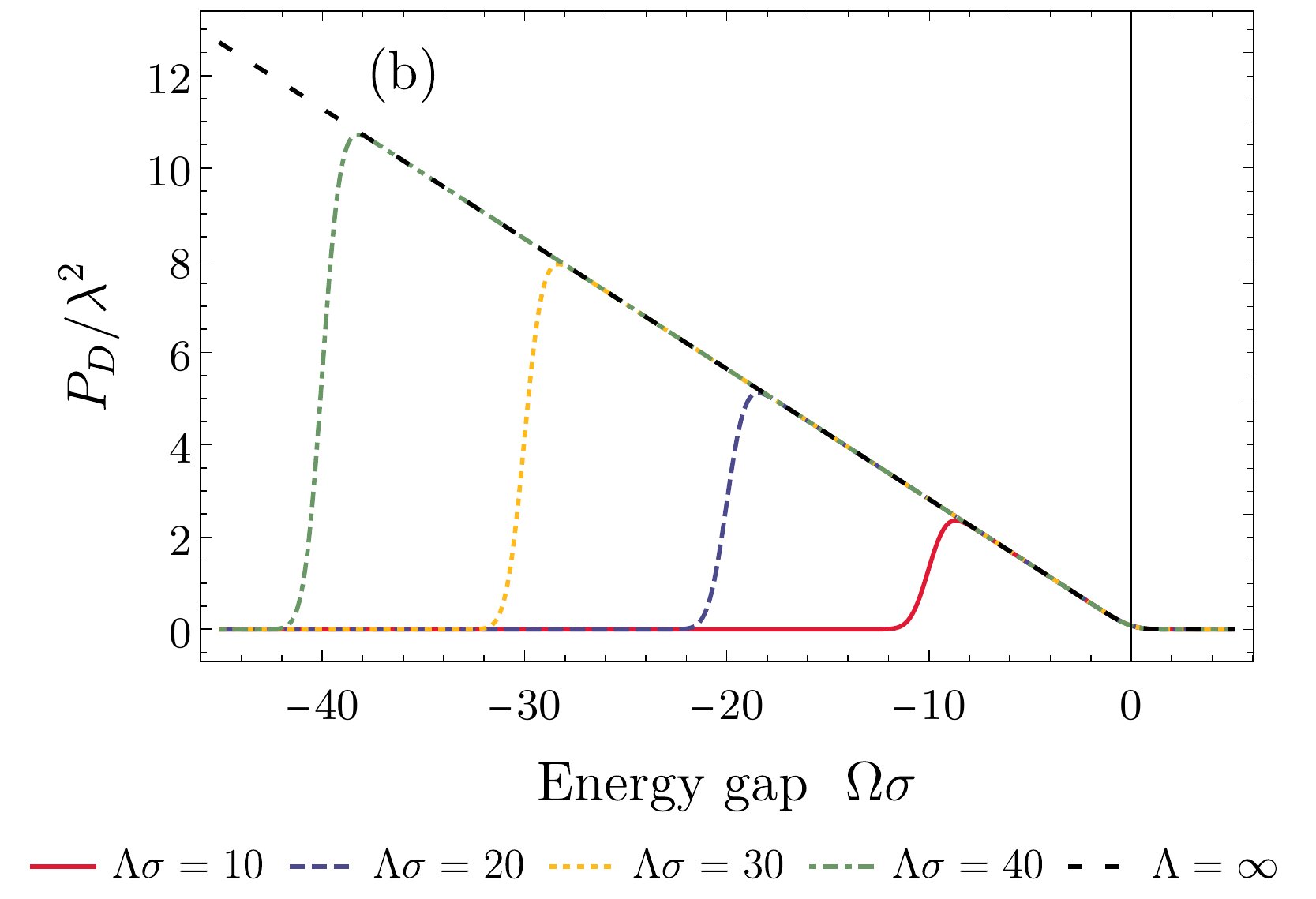}
  \caption{%
    The transition probability~$P_D$ of a single point-like detector as a function of is energy gap~$\Om$ at various values of the bandlimit~$\La$ of the field.  A negative energy gap corresponds to an initially excited detector.  When the energy gap of the detector is larger than the bandlimit, the transition probability of an initially excited detector rapidly falls off. When there is no bandlimit, the transition probability grows with energy gap.
  }
  \label{fig:PDvsOm}
\end{figure*}

For a density matrix of the form of equation \ref{eq:rhoABGeneral}, the eigenvalues of $\hat{\rh}_{AB}^{\Ga_A}$ are
\aln{
  \la_1 &= \frac{1}{2}\lb(\rh_{22}+\rh_{33}-\sqrt{(\rh_{22}-\rh_{33})^2+4\Abs{\rh_{14}}^2}\rb) \nn\\&= \frac{1}{2}\lb(P_A+P_B-\sqrt{(P_A-P_B)^2+4\Abs{X}^2}\rb) + \Or(\la^4) \nn\\
  \la_2 &= \frac{1}{2}\lb(\rh_{22}+\rh_{33}+\sqrt{(\rh_{22}-\rh_{33})^2+4\Abs{\rh_{14}}^2}\rb) \nn\\&= \frac{1}{2}\lb(P_A+P_B+\sqrt{(P_A-P_B)^2+4\Abs{X}^2}\rb) + \Or(\la^4) \nn\\
  \la_3 &= \frac{1}{2}\lb(\rh_{11}+\rh_{44}-\sqrt{(\rh_{11}-\rh_{44})^2+4\Abs{\rh_{23}}^2}\rb) \nn\\&= 0 + \Or(\la^4) \nn\\
  \la_4 &= \frac{1}{2}\lb(\rh_{11}+\rh_{44}+\sqrt{(\rh_{11}-\rh_{44})^2+4\Abs{\rh_{23}}^2}\rb) \nn\\&= 1 - P_A - P_B + \Or(\la^4)
}
The only eigenvalue that can be zero to lowest order in the coupling strength, is $\la_1$, so we calculate the negativity of the reduced density matrix (eq.\ \ref{eq:rhABPerturb}) as
\aln{
  &\mc{N}\lb(\hat{\rh}_{AB}\rb) \nn\\
  &= \max\lb[0, -\frac{1}{2}\lb(P_A+P_B-\sqrt{(P_A-P_B)^2+4\Abs{X}^2}\rb)\rb],
}
which reduces to
\aln{
  \mc{N}(\hat{\rho}_{AB}) = \max\lb[0, |X|-P_D\rb]
  \label{eq:Nsimplified}
}
when the transition probabilities of the two detectors are equal $(P_A=P_B=P_D)$.
\section{Our setup}
\label{sec:Setup}

We will now consider $(3+1)$-dimensional Minkowski space and take the two UDW detectors to be separated by a distance $S$ at rest in a common reference frame.  We will consider the detectors to be point-like
\eqn{
  F_D(\bd{x}-\bd{x}_D) = \de^{(3)}(\bd{x}-\bd{x}_D)
}
and have identical energy gaps $\Om_A=\Om_B=\Om$, and we will take the switching function to be a Gaussian with characteristic width $\si$:
\eqn{
  \ch_D(t) = \exp\lb(-\frac{t^2}{2\si^2}\rb).
}

With these assumptions the matrix elements in equations \ref{eq:PDGeneral} and \ref{eq:XGeneral} reduce to
\begin{widetext}
\aln{
  P_{A,\La}&=P_{B,\La}=P_{D,\La} = \frac{\la^2}{4\pi}\Big(\ec^{-\si^2\Om^2}-\ec^{-\si^2(\Om+\La)^2} + \sqrt{\pi}\si\Om\lb[\erf(\si\Om)-\erf\big(\si(\Om+\La)\big)\rb]\Big) \label{eq:ALambda}\\
  X_\La &= \frac{\la^2\si}{4\sqrt{\pi}S} \ec^{-\si^2\Om^2} \ec^{-S^2/(2\si)^2} \lb[\erfi\lb(\frac{S}{2\si}\rb) - \Re\lb(\erfi\lb(\frac{S}{2\si}+\iu\si\La)\rb)\rb)\rb] \nn\\
  &\qq + \iu \frac{\la^2\si^2}{2\pi S} \ec^{-\si^2\Om^2} \int_{0}^{\La} dk\ \ec^{-\si^2k^2}\erfi(\si k)\sin(Sk) \label{eq:XLambda}
}
\end{widetext}
where the subscript $\La$ is used as a reminder that the scalar field is bandlimited, $\erfi(x)\ce-\iu\erf(\iu x)$, and $\erf$ is the error function.

Using the large-$x$ expansion of
\eqn{
  \erfi(x) = \frac{2}{\sqrt{\pi}}\frac{\ec^{x^2}}{x} + \Or\lb(\frac{1}{x^3}\rb)
}
the imaginary part of $X_\La$ can be approximated as
\aln{
  \Im(X_\La) &\approx \frac{\la^2\si^2}{2\pi S}\ec^{-\si^2\Om^2} \bigg(\intallr dk\ \ec^{-\si^2k^2}\erfi(\si k)\sin(Sk) \nn\\
  &\qq - \frac{1}{\sqrt{\pi}} \int_{\La}^{\infi} dk\ \frac{\sin(Sk)}{\si k}\bigg) \nn\\
  &= \frac{\la^2\si^2}{4\pi} \frac{\ec^{-\si^2\Om^2}}{\Abs{S}} \lb(\ec^{-S^2/(2\si)^2} - 1 + \frac{2}{\pi}\SI\big(\Abs{S}\La\big)\rb)
  \label{eq:ImXSeries}
}
where $\SI(x)$ is the sine integral
\eqn{
  \SI(x) \ce \int_{0}^{x} dt \frac{\sin t}{t}
}
provided $\La\si\gg1$. Numeric testing has shown good agreement when $\La\si>5$.

\section{Transition probability}
\label{sec:TP}

First we consider the dependence of the transition probability of a single detector on its energy gap, $\Om$, and the bandlimit of the quantum field, $\La$, which is shown in figure \ref{fig:PDvsOm}.  When the detector is initially in the ground state, the transition probability falls off exponentially with increasing energy gap.  Additionally, as shown in the inset of figure \ref{fig:PDvsOm}(a), a small bandlimit lowers the transition probability, and this effect is strongest for small values of the energy gap due to the exponential suppression.  We also find that for values of the bandlimit larger than $\La\si\approx2$, the transition probability (from the ground to the excited state) is nearly indistinguishable from the non-bandlimited case regardless of the energy gap.

In order to better understand this behaviour, it is useful to write the transition probability in terms of the non-bandlimited case as
\eqn{
  P_{D,\La}(\Om) = P_{D,\infi}(\Om) - P_{D,\infi}(\Om+\La) - \frac{\la^2\si\La}{4\sqrt{\pi}}\erfc\big(\si(\Om+\La)\big)
}
where $\erfc(x) \ce 1-\erf(x)$. Since the bandlimit enters the expression like a modified energy gap, it is easy to see that for large values, both $P_{D,\infi}(\Om+\La)$ and $\erfc\big(\si(\Om+\La)\big)$ are exponentially suppressed and ${P_{D,\La}(\Om) \approx P_{D,\infi}(\Om)}$.

If the detector was initially excited, which is mathematically equivalent to taking $\Om<0$, the transition probability increases linearly with increasing energy gap up to a maximum value after which it exponentially falls off, which is seen in the negative-$\Omega$ regions of figure \ref{fig:PDvsOm}.  As the bandlimit is increased, we find that the maximum value of de-exciation probability occurs at larger (more-negative) values of energy gap and that the falloff after is sharper.

We define $\Om_\tx{crit}$ to be the value of energy gap associated with the maximum de-excitiation probability.  Since the transition probability falls off so rapidly when $\Abs{\Om}<\Abs{\Om_\tx{crit}}$, this energy gap can be interpreted as the largest energy gap for which exited detectors are able to decay into the bandlimited field (any larger and the de-excitiation probability is exponentially suppressed).  This is not surprising, since bandlimiting the field also puts limits on the frequency of modes that exist in the theory, and since we have only calculated the reduced density matrix to lowest order in the coupling constant, we only consider single-mode de-excitiations.  In other words there is no room in the field for the detector to decay into.

\begin{figure}[ht]
  \centering
  \includegraphics[width=\linewidth]{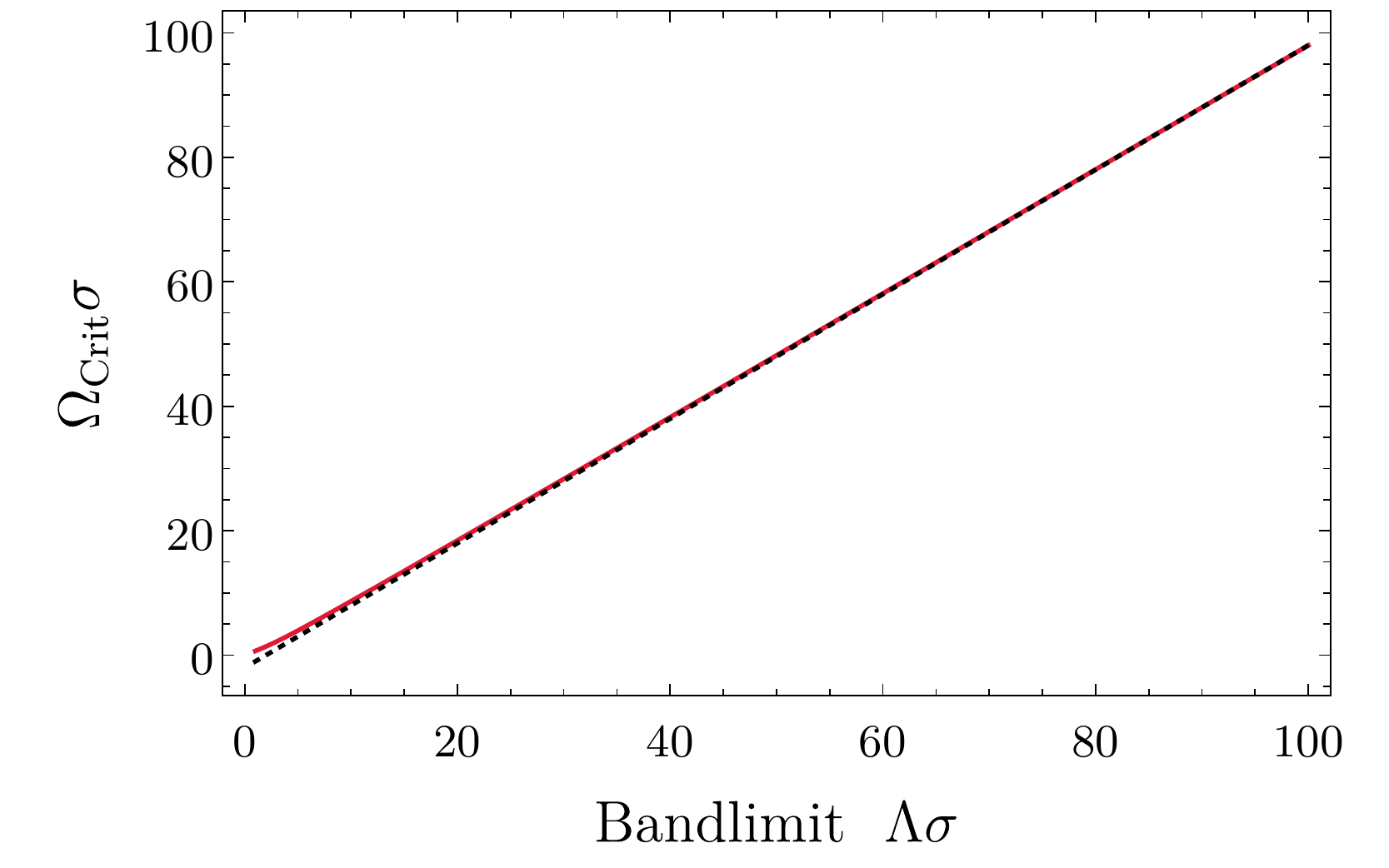}
  \caption{%
    The energy gap of the detector corresponding to the maximum de-excitation probability as a function of the bandlimit (red solid line).  The de-excitation probability rapidly falls off for detectors with energy gaps greater than this critical value.  The black dotted line corresponds to a line of best fit $\Omega_{\text{Crit}} = \Lambda - 2/\sigma$.
  }
  \label{fig:OmegaCrit}
\end{figure}

The dependence of the $\Om_\tx{crit}$ on the bandlimit of the scalar field is plotted in figure \ref{fig:OmegaCrit}, where the detector is assumed to be initially excited.  $\Om_\tx{crit}$ grows with nearly the same rate as the bandlimit of the field.  The most discrepancy occurs at small values of $\Om\si$ where there is a high degree of quantum uncertainty in the effective energy gap of the detector.  At large values of $\Om_\tx{crit}$ and $\La$, the relationship approaches the linear relationship $\Om_\tx{crit} = \La-2/\si$.

\section{Negativity}
\label{sec:EH}

Now we consider the entanglement between two identical detectors $A$ and $B$, which are initialised in the ground state and separated by a distance $S$.  We first note that the qualitative dependence of negativity on the energy gap of the detectors does not change significantly as the bandlimit is changed.  The actual value of the negativity at given value of $\Om$ is much lower than the non-bandlimited case for a small enough value of $\La$, and as $\La$ increases, the negativity oscillates around the non-bandlimited value.  For this reason we only consider detectors with an energy gap of $\Om\si=0.01$ throughout this section.
\begin{figure}[t]
  \centering
  \includegraphics[width=\linewidth]{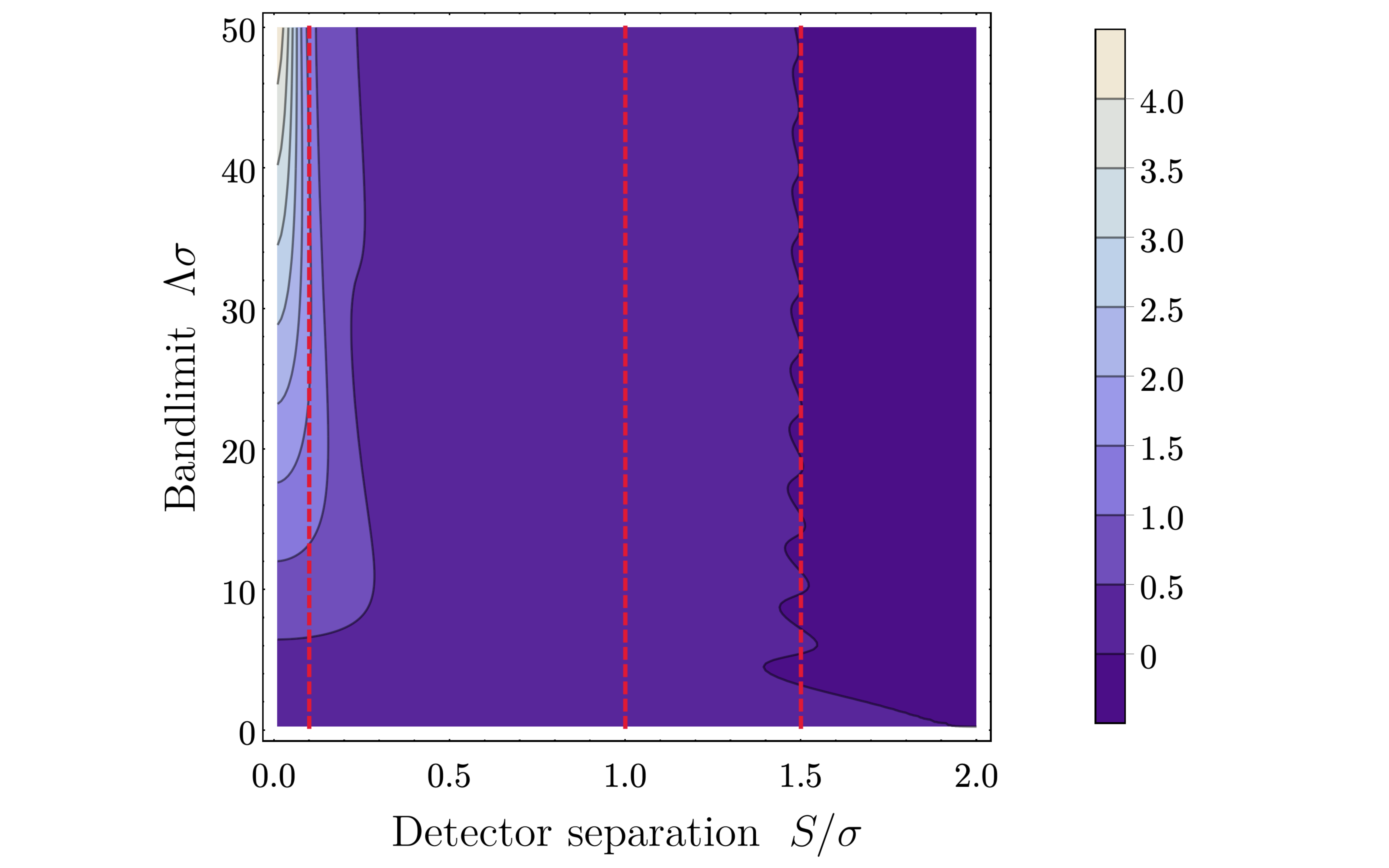}
  \caption{%
    The negativity of the reduced density matrix for the detectors' internal states, $\mathcal{N}(\hat{\rho}_{AB})/\lambda^2$, as a function of detector separation~$S$ and the bandlimit~$\Lambda$ of the field when the detectors have an energy gap of~$\Omega\sigma=0.01$.  The darkest region corresponds to a value of zero negativity when the two detectors remain in a separable state.  The red dashed line mark the constant separation slices shown in figure \ref{fig:NvsLambda}.
  }
  \label{fig:NContour}
\end{figure}
\begin{figure*}[t]
  \centering
  \includegraphics[trim=25 0 25 0, clip, width=0.32\linewidth]{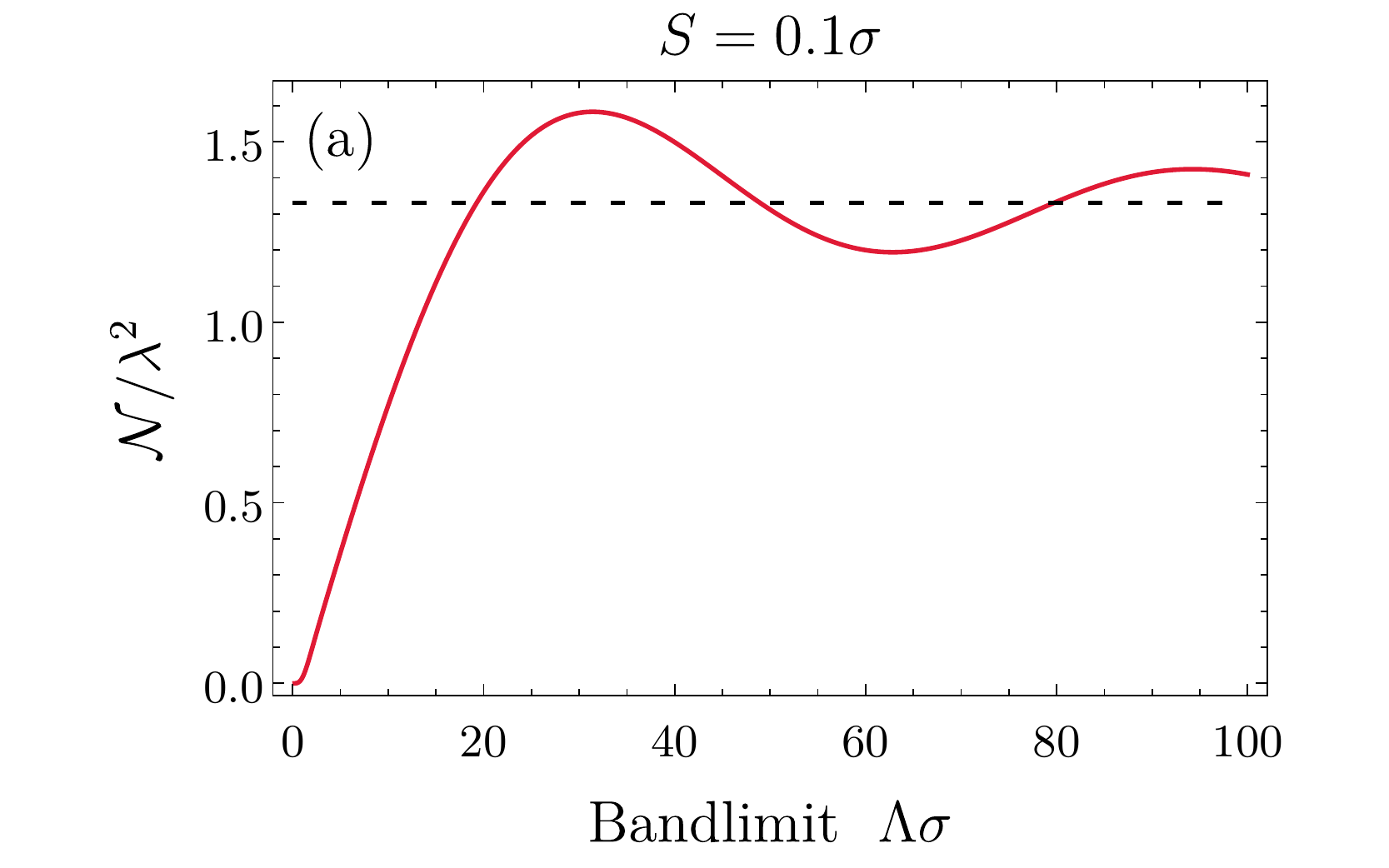}
  \includegraphics[trim=25 0 25 0, clip, width=0.32\linewidth]{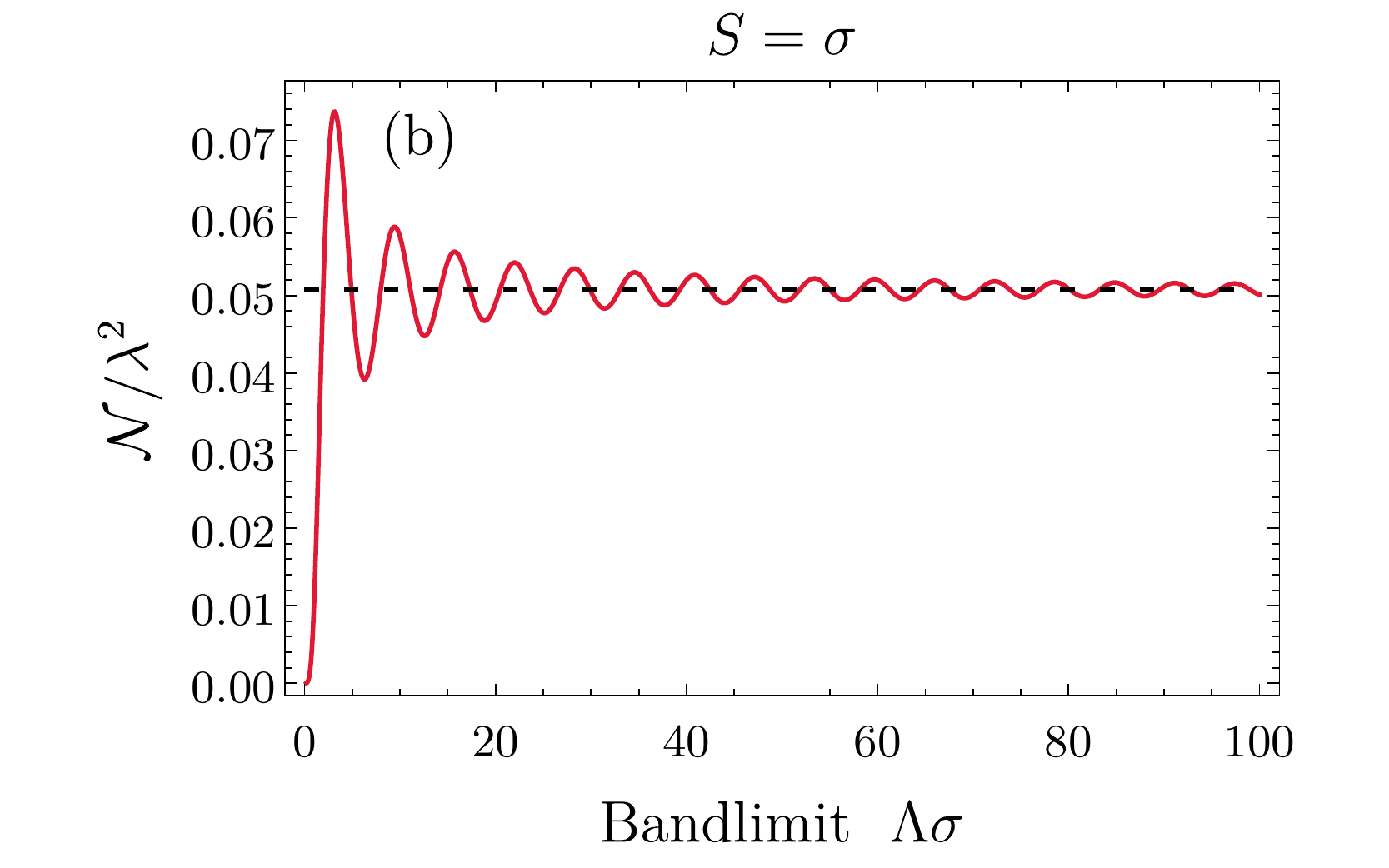}
  \includegraphics[trim=25 0 25 0, clip, width=0.32\linewidth]{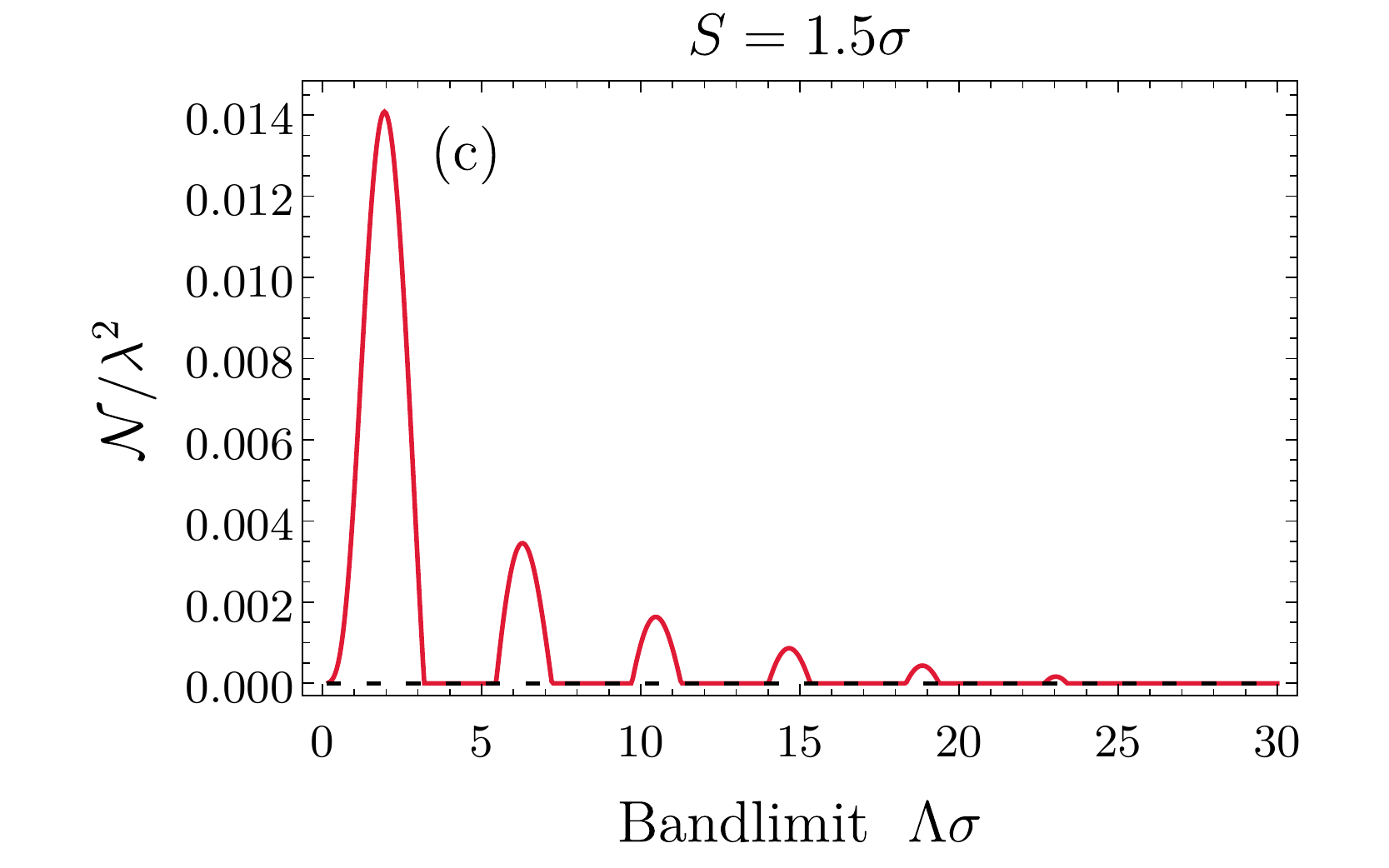}
  \\
  \includegraphics[width=0.65\linewidth]{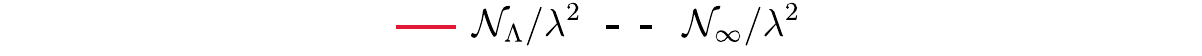}
  \caption{%
    Red solid line: The negativity of the reduced density matrix, $\mathcal{N}(\hat{\rho}_{AB})/\lambda^2$, as a function of the bandlimit~$\Lambda$. Black dashed line: The value of the negativity when there is no bandlimit $(\Lambda\to\infty)$.  The detector energy gap is set to $\Omega\sigma=0.01$.
  }
  \label{fig:NvsLambda}
\end{figure*}

In figure \ref{fig:NContour} we plot the dependence of the entanglement harvested on the bandlimit and the separation of the detectors.  There are striking oscillations in the contours when considering a fixed value of separation, which we plot in figure \ref{fig:NvsLambda}.  The frequency of these oscillations decreases when the detector separation is decreased.  We also note the negativity decreases with increasing detector separation at fixed values of the bandlimit.

We take a closer look at the oscillations in negativity with respect to the bandlimit at fixed values of detector separation in figure \ref{fig:NvsLambda}.  For values of separation where entanglement harvesting possible in the non-bandlimted case (figures \ref{fig:NvsLambda}(a) and \ref{fig:NvsLambda}(b)), we find the negativity increases with increasing values of the bandlimit up to some maximum value.  As the bandlimit increases further, the negativity will exhibit damped oscillations around the non-bandlimied value.  The frequency of these oscillations increases as the detector separation is increased.

These oscillations can also be present when separation of the detectors is too large for entanglement harvesting to be possible in the non-bandlimited case, which we show in figure \ref{fig:NvsLambda}(c).  In this case, the damped oscillations that are still present in the function $|X|-P_D$ are cut off for negative values---but remain when $|X|>P_D$.  This results in values of the bandlimit where entanglement harvesting is possible, even if it not possible when there is no bandlimit.
\begin{figure}[t]
  \centering
  \includegraphics[width=\linewidth]{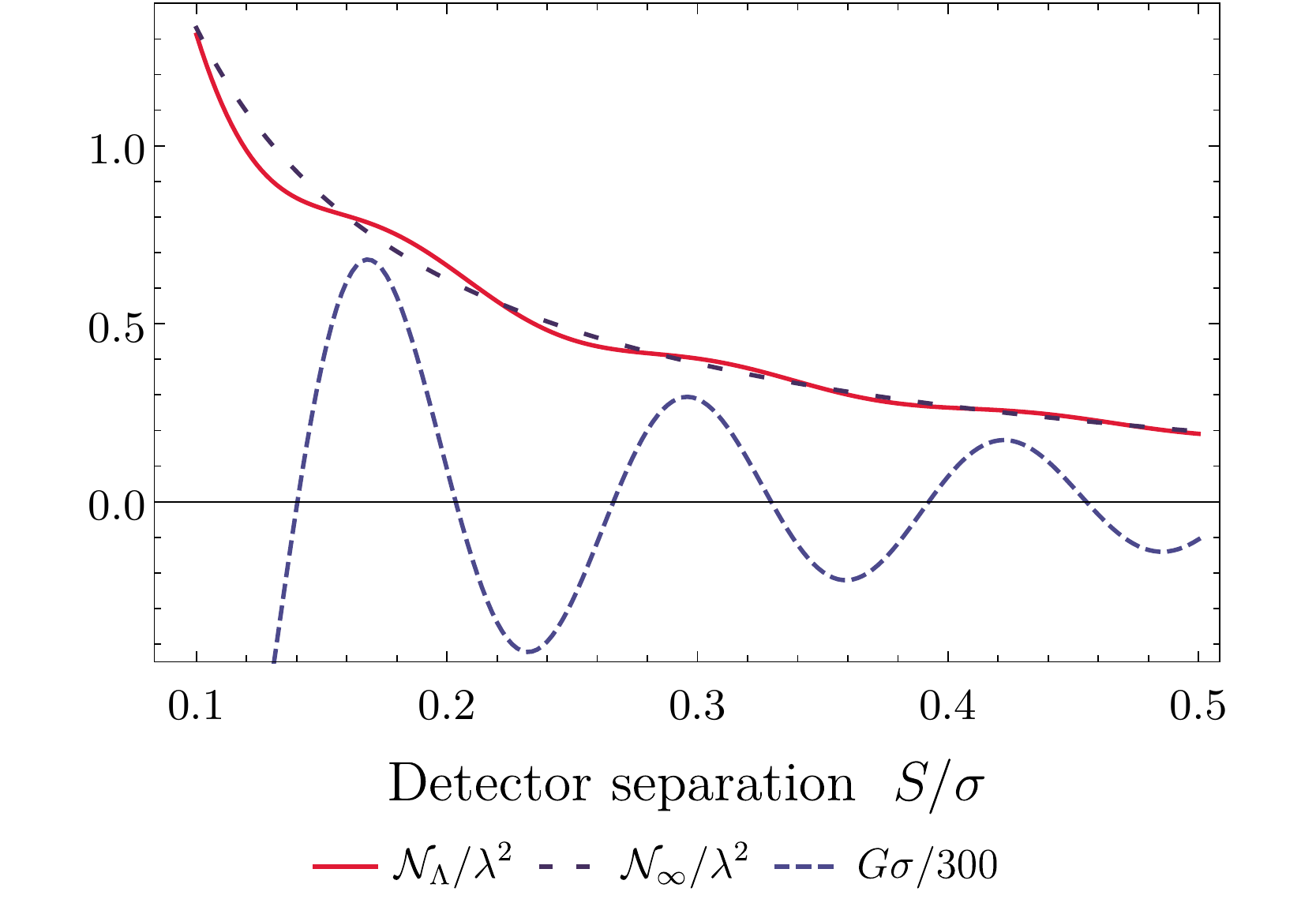}
  \caption{%
    Red solid line: The negativity of the reduced density matrix, $\mathcal{N}(\hat{\rho}_{AB})/\lambda^2$, as a function of the detector separation~$S$ when the bandlimit is set to $\La\si=50$.  Black dashed line: The negativity as a function of detector separation when there is no bandlimit $(\La\to\infi)$. Blue dashed line: A plot of the effective spatial profile $G = \sqrt{\frac{2}{\pi}}\La^2\frac{j_1(\La S)}{S}$ for $\La\si=50$ scaled by a factor of 1/300 for ease of plotting.  The energy gap is set to $\Om\si=0.01$.  The overall decay of the bandlimited plot matches the decay of the non-bandlimed one, and oscillations in the bandlimited plot match up with the oscillations in the effective spatial profile.
  }
  \label{fig:NvsS}
\end{figure}

In figure \ref{fig:NvsS}, we plot the negativity as a function of detector separation at a fixed bandlimit of $\La=50\si$ and $\La\to\infi$.  Similar to the plots in figure \ref{fig:NvsLambda}, we find the negativity exhibits damped oscillations around the non-bandlimited curve.  Interestingly, we also find that these oscillations match up with the oscillations on the effective spatial profile (equation \ref{eq:thinc}), which implies that the oscillations in negativity may be a result of increased/decreased overlap of the detectors' effective spatial profiles.

\begin{figure}[t]
  \centering
  \includegraphics[width=\linewidth]{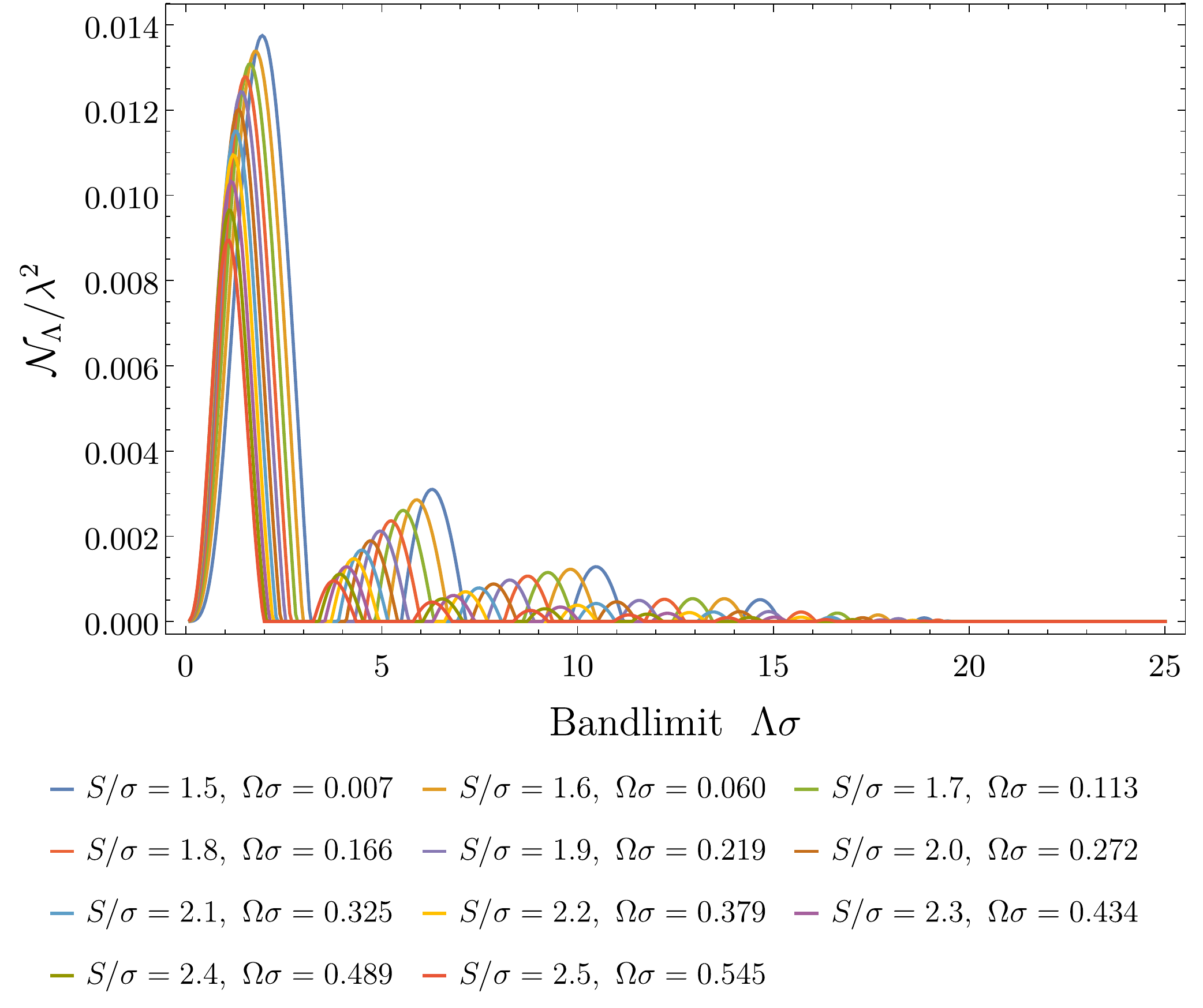}
  \caption{%
    An array of detectors with properly tuned energy gaps are set at different separation distances so that equation~\ref{eq:BuildTheArray} is satisfied at $\La_\tx{threshold}\si=20$.  If the negativity of any one of the pairs of detectors (each corresponding to a different experiment) has nonzero negativity, then $\La<\La_\tx{threshold}$.
  }
  \label{fig:Array}	
\end{figure}

Since neither the transition probability of the detectors nor the real part of the matrix element $X_\La$ depend significantly on the bandlimit when it is large, the oscillatory behaviour in both figures \ref{fig:NvsLambda} and \ref{fig:NvsS} must be due the imaginary part of $X$, and can be approximated as resulting from the $\SI(S\La)$ term in equation \ref{eq:ImXSeries}.  Therefore, the oscillations with respect to $\La$ in the imaginary part of $X$ can be shown to be upper bounded by
\eqn{
  \frac{\la^2\si}{2\pi^{3/2}} \frac{\ec^{-\si^2\Om^2}}{S^2\La} + \Im(X_\infi).
  \label{eq:Envelope}
}
Using this envelope, it is possible to tune the separation of pair of detectors of fixed energy gap so that the entanglement between them is zero for all values of the bandlimit greater than some chosen threshold and nonzero for \textit{some} values of the bandlimit less than that threshold.  In other words, for a chosen value of $\La_\tx{threshold}$ and a given value of $\Om$ (or $S$), find a value of  $S$ (or $\Om$) so that
\aln{
  &P_{D,\La_\tx{threshold}} = \nn\\ &%
  \Bigg|\Re\lb(X_{\La_\tx{threshold}}\rb) + \iu\lb(\frac{\la^2\si}{2\pi^{3/2}} \frac{\ec^{-\si^2\Om^2}}{S^2\La_\tx{threshold}} + \Im(X_{\infi})\rb)\Bigg|
  \label{eq:BuildTheArray}
}
is satisfied.
It is important to note that if only one pair is used, then there will be regions of zero negativity at some values of $\La < \La_\tx{threshold}$. For example, consider a value of $\La\si=4$ in figure \ref{fig:NvsLambda}(c). However, it is possible to build up an array of such pairs of detectors, each at different values of separation, so that the frequency of oscillations in negativity is different.  This can be used to fill in the gaps of each individual curve and in principle test if a quantum field has a bandlimit $\La\le\La_\tx{threshold}$.  An example of such an array is shown in figure~\ref{fig:Array}.

\section{Going beyond perturbative: $\de$-Switching}
\label{sec:delta}

An alternative approach to using perturbation theory to compute the reduced density matrix of the two detectors is to implement $\de$-switching
\eqn{
  \ch_D(t)=\sqrt{2\pi}\si\de(t-T_D)
}
which can be related to the switching used in the previous section by defining the Dirac delta as the weak limit of a normalised Gaussian,
\eqn{
  \de(x) = \lim_{\si\to0} \frac{\ec^{-x^2/(2\si^2)}}{\sqrt{2\pi}\si}.
  \label{eq:DeltaLimit}
}
This choice of switching allows for the calculation of the time evolution operator (eq.\ \ref{eq:TimeEvOp}) exactly.  Following the work of Simidzija and Mart\'in-Mart\'inez \cite{Simidzija:2017kty}, we define detector $A$ to be the one that switches first $(T_A \le T_B)$ and calculate the time-evolution operator as
\aln{
  \hat{U}_{\de} &= \lb(\id_A\tp\id_B\tp\cosh(\hat{Y}_B) + \id_A\tp\hat{\mu}_B(T_B)\tp\sinh(\hat{Y}_B)\rb) \nn\\
  &\times \lb(\id_A \tp \id _B \tp \cosh(\hat{Y}_A) + \hat{\mu}_A(T_A)\tp\id_B\tp\sinh(\hat{Y}_A)\rb)
  \label{eq:TimeEvOpdelta}
}
where
\eqn{
  \hat{\mu}_D(t) = \ec^{\iu\Om_D\ta_D(t)}\sipl_D+\ec^{-\iu\Om_D\ta_D(t)}\simi_D
}
and
\eqn{
  \hat{Y}_D \ce -\iu\la\sqrt{2\pi}\si \lb.\frac{d\ta_D}{dt}\rb|_{t=T_D} \int d^n\bd{x} F_D(\bd{x}-\bd{x}_D)\hat{\ph}(\bd{x},T_D).
}

The detector-field system is initially in the state $\Ket{\Ps_0} = \Ket{0}_A\tp\Ket{0}_B\tp\Ket{0}_\ph$ and the final state of the two detector subsystem is
\eqn{
  \hat{\rh}_{AB} = \Tr_\ph\lb[\hat{U}_\de\Kb{\Ps_0}{\Ps_0}\hat{U}_\de^\ct\rb] = \pmx{
    \rh_{11} & 0 & 0 & \rh_{14} \\
    0 & \rh_{22} & \rh_{23} & 0 \\
    0 & \rh_{23}^* & \rh_{33} & 0 \\
    \rh_{14}^* & 0 & 0 & \rh_{44}
  }
}
where
\subeqn{
  \aln{
    \rh_{11} &= \frac{1}{4} \Big(1+f_A+f_B\cos(2\tht)+f_Af_B\cosh(\om)\Big) \\
    \rh_{14} &= \frac{1}{4} \ec^{-\iu(\Om_AT_A+\Om_BT_B)} f_B \Big(\iu\sin(2\tht)+f_A\sinh(\om)\Big) \\
    \rh_{22} &= \frac{1}{4} \Big(1+f_A-f_B\cos(2\tht)-f_Af_B\cosh(\om)\Big) \\
    \rh_{23} &= -\frac{1}{4} \ec^{-\iu(\Om_AT_A-\Om_BT_B)} f_B\Big(\iu \sin(2\tht)+f_A\sinh(\om)\Big) \\
    \rh_{33} &= \frac{1}{4} \Big(1-f_A+f_B\cos(2\tht) - f_Af_B\cosh(\om)\Big) \\
    \rh_{44} &= \frac{1}{4} \Big(1-f_A-f_B\cos(2\tht)+f_Af_B\cosh(\om)\Big)
  }
}
and we have defined
\begin{figure*}[t]
  \centering
  \includegraphics[width=0.49\linewidth]{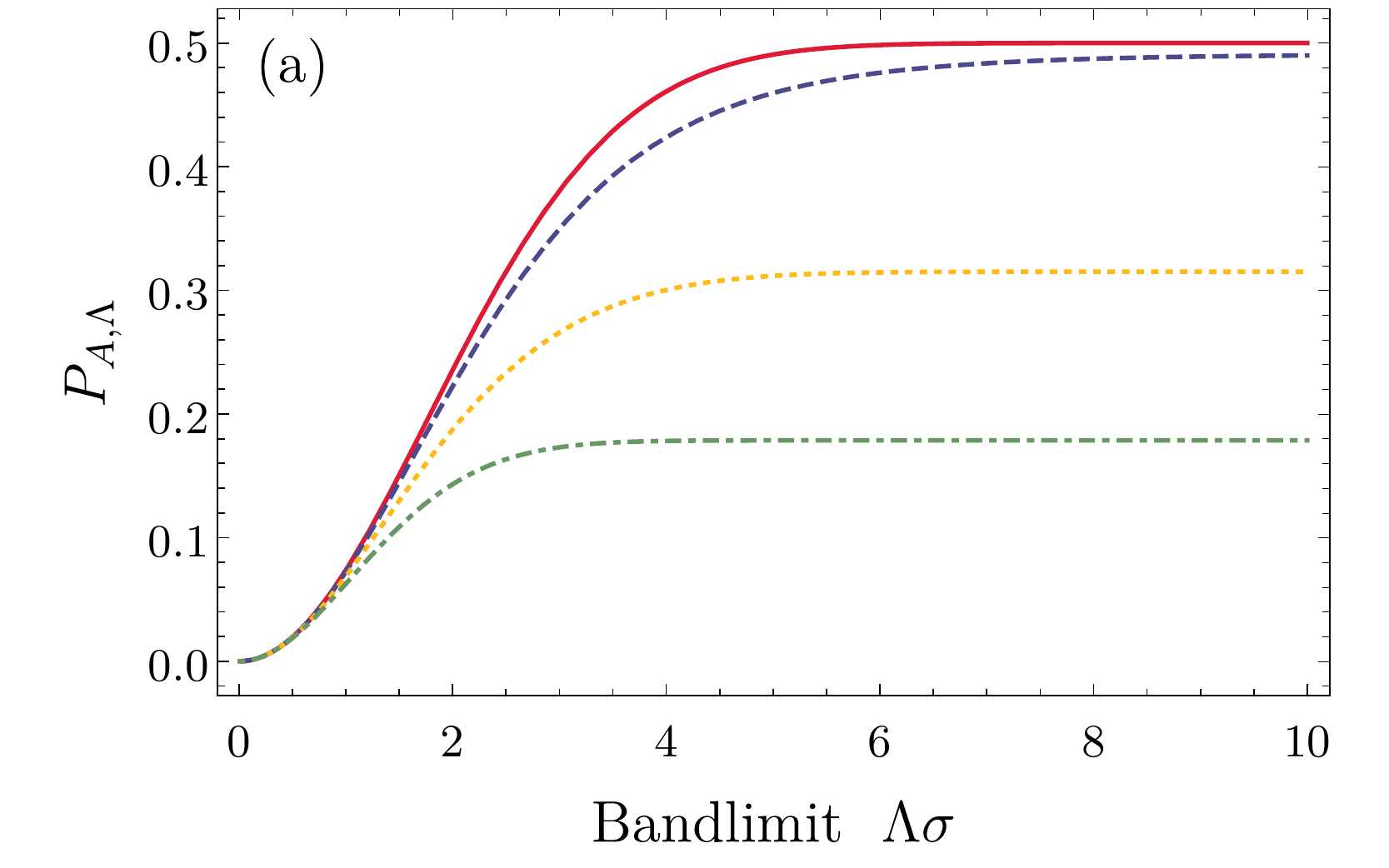}
  \includegraphics[width=0.49\linewidth]{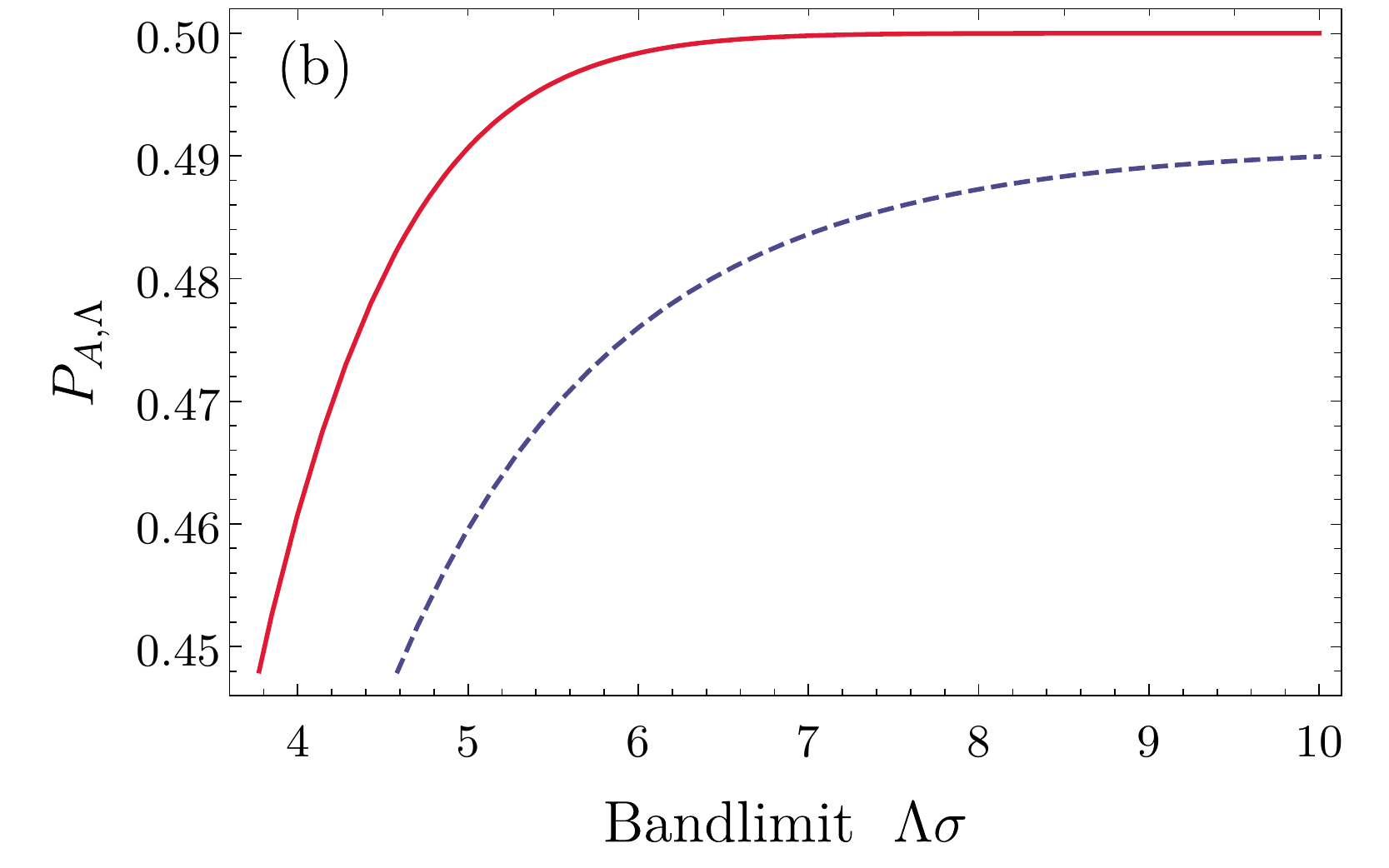}
  \\
  \includegraphics[width=0.65\linewidth]{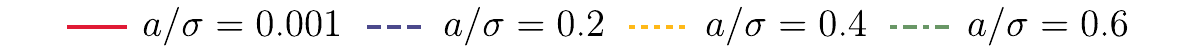}
  \caption{%
    The transition probability of a single detector interacting with a bandlimited scalar field with delta switching as a function of the bandlimit $\Lambda$ for various widths of the spatial profile.  As the bandlimit increases, the transition probability quickly asymptotes to the non-bandlimited $(\La\to\infi)$ value.  The right hand figure details the asymptotic behaviour of the narrow (${a=0.001\si}$ and ${a=0.2\si}$) detectors.  The interaction strength is set to $\la=1$.
  }
\label{fig:deltaPA}
\end{figure*}
\begin{widetext}
\subeqn{
  \aln{
    f_D &\ce \exp\lb(-2\pi\la^2\si^2 \int_{\Abs{\bd{k}}<\La} \frac{d^n\bd{k}}{\Abs{\bd{k}}}\Abs{\ti{F}_D(\bd{k})}^2\rb) \label{eq:fD}\\
    \tht &\ce -\iu\lb[\hat{Y}_A,\hat{Y}_B\rb] = -\iu\pi\la^2\si^2 \int_{\Abs{\bd{k}}<\La} \frac{d^n\bd{k}}{\Abs{\bd{k}}} \Big(\ti{F}^*_A(\bd{k})\ti{F}_B(\bd{k}) \ec^{\iu\Abs{\bd{k}}(T_A-T_B)} \ec^{-\iu\bd{k}\cdot(\bd{x}_A-\bd{x}_B)} - \tx{H.c.}\Big)\\
    \om &\ce 2\pi\la^2\si^2 \int_{\Abs{\bd{k}}<\La} \frac{d^n\bd{k}}{\Abs{\bd{k}}} \Big(\ti{F}_A^*(\bd{k})\ti{F}_B(\bd{k})\ec^{\iu\Abs{\bd{k}}(T_A-T_B)} \ec^{-\iu\bd{k}\cdot(\bd{x}_A-\bd{x}_B)} + \tx{H.c.}\Big).
  }
}
In order to compare to our perturbative results, we again consider $(3+1)$-dimensional Minkowski space and take the UDW detectors to be identical $(\Om_A=\Om_B=\Om)$ and at rest in a common reference frame and separated by a distance $S$.  To avoid divergences in $\tht$ and $\om$ in the non-bandlimited limit, we set the spatial profile of the detectors to be Gaussian
\eqn{
  F_D(\bd{x}) = \frac{1}{(2\pi)^{3/2}a^3}\exp\lb(-\frac{\Abs{\bd{x}}^2}{2a^2}\rb)
  \label{eq:GaussianProfile}
}
rather than point like.  With these simplifications, the matrix-element functions become
\subeqn{
  \aln{
    f_A&=f_B=f= \exp\lb(-\frac{\la^2\si^2}{2\pi a^2}\lb(1-\ec^{-a^2\La^2}\rb)\rb) \label{eq:fDwithProfile}\\
    \tht & = \frac{\la^2\si^2}{4\sqrt{\pi}aS}\sum_{j\in\{0,1\}} (-1)^j\exp\lb(-\frac{\big(S+(-1)^jT\big)^2}{4a^2}\rb) \Im\lb[\erfi\lb(\frac{S+(-1)^jT}{2a}+\iu a\La\rb)\rb] \\
    \om &= -\frac{\la^2\si^2}{2\sqrt{\pi}aS} \sum_{j\in\{0,1\}} \exp\lb(-\frac{\big(S+(-1)^jT\big)^2}{4a^2}\rb) \Bigg[\erfi\lb(\frac{S+(-1)^jT}{2a}\rb)  - \Re\lb(\erfi\lb(\frac{S+(-1)^jT}{2a}+\iu a\La\rb)\rb)\Bigg]
  }
}
where $T \ce T_B-T_A$.
\end{widetext}

Finally, we calculate the transition probability of the detectors to be
\aln{
  &P_{A,\La} = \frac{1}{2}(1-f), &P_{B,\La}=\frac{1}{2}\big(1-f\cos(2\tht)\big),
  \label{eq:deltaPD}
}
which are only equal when the field commutator vanishes.  Since the time evolution operator in equation \ref{eq:TimeEvOpdelta} is calculated exactly rather than through a Dyson series expansion, this model allows for the backreaction of the detectors on the field.  If detector $A$ is in causal contact with detector $B$, then $B$ will be interacting with the field that was modified by $A$.  It is also worth noting that the transition probabilities are independent of the energy gap of the detector, and so they can be interpreted as both excitation and de-excitation probabilities.

\subsection{One detector}

First, we consider the delta-switching interaction of a single detector with a bandlimited scalar field.  Since the bandlimit acts as an effective spatial profile, we can safely take the $a\to0$ limit of the equation \ref{eq:GaussianProfile} and again consider a point-like detector.  In this limit, the transition probability is
\eqn{
  \lim_{a\to0}P_{A,\La} = \frac{1}{2}\lb[1-\exp\lb(-\frac{\la^2\si^2\La^2}{2\pi}\rb)\rb].
}
Due to the exponential decay of $f$ with increasing $\La$, the transition probability of a single point-like detector interacting with a non-bandlimited field becomes
\eqn{
  \lim_{\La\to\infi}\frac{1}{2}\lb[1-\exp\lb(-\frac{\la^2\si^2\La^2}{2\pi}\rb)\rb] = \frac{1}{2}.
}
Additionally, this limit is well defined since taking the limits in the other order (first $\La\to\infi$ then $a\to 0$) produces the same result.

One does not need to take limits of equation \ref{eq:fDwithProfile} to calculate the transition probability of a point-like $(\ti{F}(\bd{k})=(2\pi)^{-n/2}$) detector in a non-bandlimied field. Instead, one can directly consider equation \ref{eq:fD}, which leads to
\eqn{
  P_{A,\infi} = \frac{1}{2}\lb[1-\exp\lb(-\frac{2\pi\la^2\si^2}{(2\pi)^n} \int \frac{d^n\bd{k}}{\Abs{\bd{k}}}\rb)\rb] = \frac{1}{2}
  \label{eq:ActualAllDelta}
}
provided $n\ge 2$.  In other words, if a two level detector is initialised in the ground (or excited) state and is coupled to the field at a single point in space and time, the result would be the completely mixed state. %

This is still consistent with the conventional idea that the transition probability of point-like detector with delta switching will diverge \cite{Pozas-Kerstjens:2017xjr} if one considers the perturbative calculation, since to lowest order in the coupling strength
\eqn{
  P_{A,\infi} \approx \frac{2\pi\la^2\si^2}{(2\pi)^n} \int \frac{d^n\bd{k}}{\Abs{\bd{k}}} \to \infi.
}
This divergence is solely a mathematical artefact of using perturbation theory, however. The physical behaviour in this case (interaction at a single spacetime point) is in fact well defined, finite, and given by equation~\ref{eq:ActualAllDelta}.

In figure \ref{fig:deltaPA}, we plot the transition probability of a single detector---with various widths of its Gaussian spatial profile, interacting via delta switching---versus the bandlimit of the scalar field.  Since there is no dependence on the energy gap of the detector, this can be interpreted as either the probability of excitation or de-exitiation. Regardless of the width of the detector, the transition probability increases with increasing value of the bandlimit and asymptotes to the non-bandlimited value.

We make comparisons to section \ref{sec:TP} by considering the small-$\sigma$ limit where the unitless energy gap~$\Om\si$ will also be close to zero.  In this regime, the dependence of the transition probability on the bandlimit can be best seen in the inset of figure \ref{fig:PDvsOm}(a).  Here, the same behaviour is seen: when the bandlimit is small, the transition probability of the detector is increased when the bandlimit is increased. But for larger values of the bandlimit, the transition probability is indistinguishable from the non-bandlimtied $(\La\to\infi)$ case.

The width of the detector determines how quickly the transition probability approaches its corresponding non-bandlimed value, which can be seen in figure \ref{fig:deltaPA}.  When the spatial profile of the detector is relativity wide and the width is increased, the transition probability flattens out and approaches its asymptotic value at lower values of the bandlimit.  However, when the spatial profile is very narrow, as seen in figure \ref{fig:deltaPA}(b), the transition probability flattens for the detector with a width of $a=0.001\si$ at a smaller value of the bandlimit than the detector with a width of $a=0.2\si$.

\begin{figure}[t]
  \centering
  \includegraphics[width=\linewidth]{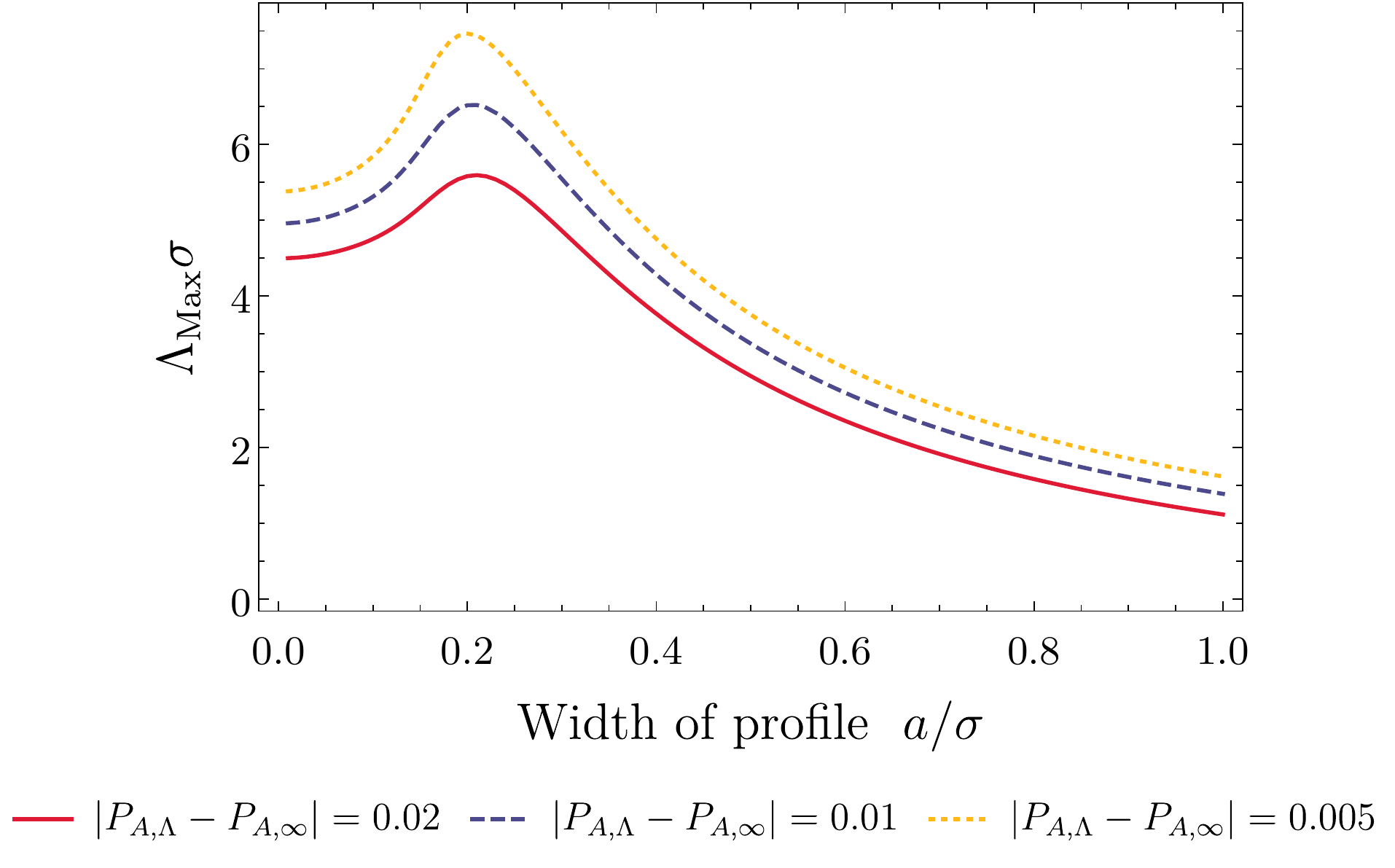}
  \caption{%
    The value of the maximum bandlimit~$\La_{\text{Max}}$ such that the difference between the transition probability in the bandlimited field and non-bandlimited field $\big(\Abs{P_{A,\La_\tx{Max}}-P_{A,\infi}})$ is equal to a specified (arbitrary) tolerance (0.02, 0.01, 0.005) as a function of the width of the spatial profile of the detector.  When the bandlimit is larger than $\La_{\tx{Max}}$, the difference will be less than the tolerance.  For all three values, a maximum occurs near at width of $a\approx0.2\si$, indicting the transition probability of a detector with this width is more sensitive to the bandlimit than a wider or narrower detector.  The interaction strength is set to $\la=1$.
  }
  \label{fig:LambdaMax}
\end{figure}

This observation is made more precise in figure \ref{fig:LambdaMax}, where we plot the value of the bandlimit, denoted $\La_\tx{Max}$, for which the absolute difference between the transition probability and its corresponding non-bandlimited value is equal to a chosen tolerance as a function of the width of the spatial profile. This quantity was chosen for two reasons.  First, it provides a concrete measure of determining that the transition probability is ``close'' to its asymptotic value, and second, from an operational perspective the tolerance can be interpreted as the resolution of some experiment where UDW detectors are used to determine the bandlimit of the field.

We find that for all three values of the chosen tolerance, the behaviour of the absolute difference is the same.  When the spatial profile is very narrow, increasing the width of the profile actually increases the value of $\La_{\max}$ up to a maximum value, which occurs at $a\approx0.2\si$.  As the width is increased further, the value of $\La_\tx{Max}$ rapidly decreases.

The region where increasing the width decreases $\La_\tx{Max}$ is easily explained by naively looking at equation \ref{eq:NewProfile}.  When the bandlimit is large enough such that $\frac{1}{\La} \ll a$, the effective profile, %
\aln{
  G(\bd{x}) &= \frac{\ec^{-a^2\La^2/2}}{(2\pi)^{3/2} a^3} \Bigg[\ec^{(a^4\La^2-\Abs{\bd{x}}^2)/(2a^2)} \Re\lb(\erf\lb(\frac{a^2\La+\iu\Abs{\bd{x}}}{\sqrt{2}a}\rb)\rb) \nn\\ &\qquad - \frac{\sqrt{2}a\La}{\sqrt{\pi}} \sinc\big(\La \Abs{\bd{x}}\big)\Bigg],
}
will be dominated by the spatial profile of the detector, and the transition probability will be nearly the same as the non-bandlimited case.  However, when the width of the spatial profile is very small, the complicated relationship between the width and the bandlimit in the effective spatial profile leads to the increase in sensitivity with increasing width.  The takeaway message is this: When comparing detectors with a Gaussian profile and delta switching, the detector most sensitive to the bandlimit is not the one with the narrowest profile, as one may naively think, but rather one with a width of approximately $a\approx0.2\si$

\begin{figure*}[t]
  \centering
  \includegraphics[width=0.49\linewidth]{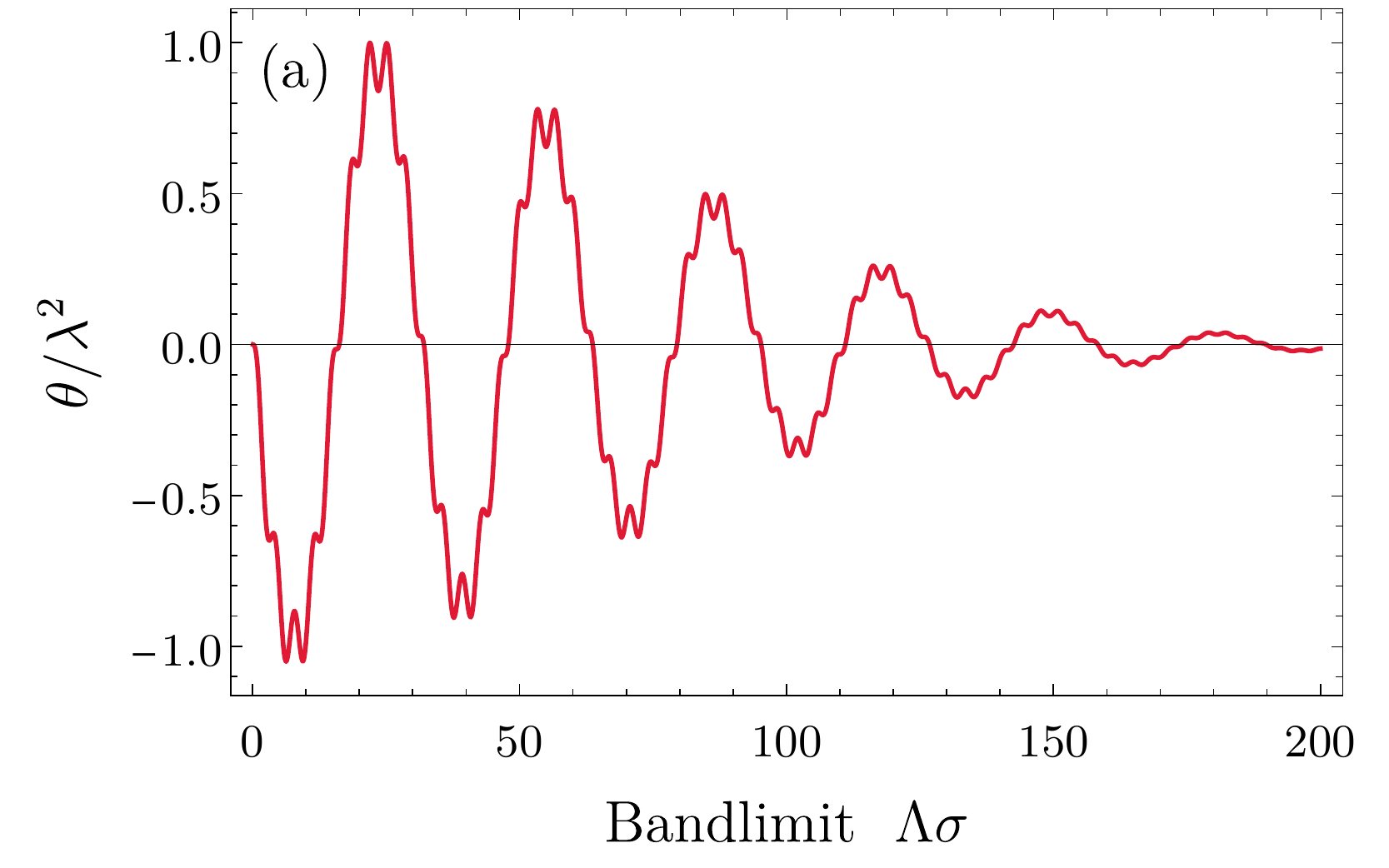}
  \includegraphics[width=0.49\linewidth]{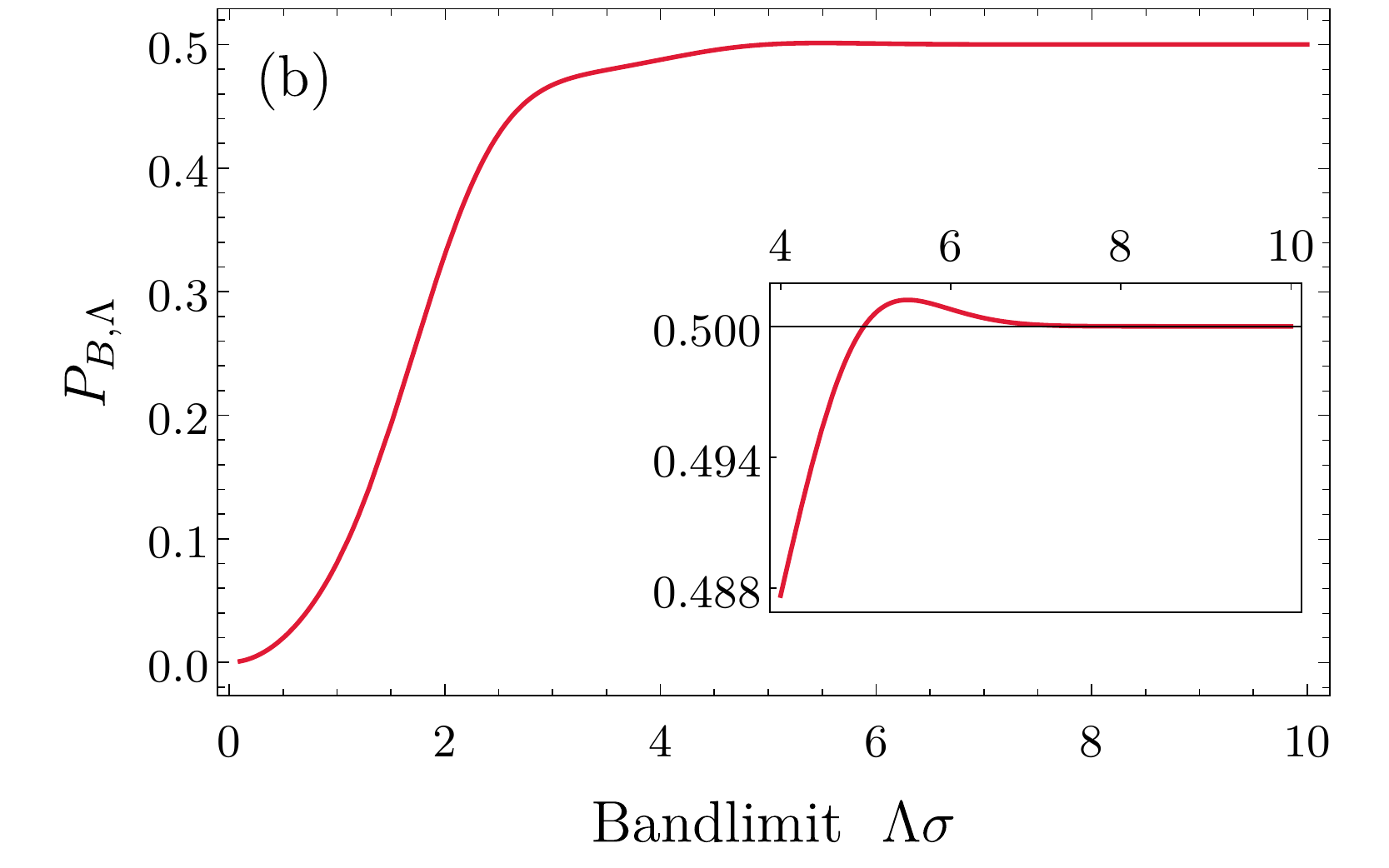}
  \caption{%
    A plot of (a) the parameter $\tht$, which is proportional to the effective field commutator between detectors $A$ and $B$, and (b) the transition probability of detector $B$ as a function of the bandlimit~$\La$.  The negative values in the parameter $\tht$ result in values of $P_B>\frac{1}{2}$. Although $\tht$ shows dependency on the bandlimit over a large range, we find that $P_{B,\Lambda}$ is only sensitive to the bandlimit when it is small ($\Lambda<7\sigma$).  The width of the spatial profile is set to $a=0.01\si$, the detectors have a time delay of $T=\si$ and a separation of $S=0.8\si$, and the interaction strength is set to $\la=1$.
  }
  \label{fig:deltaThetaAndPB}
\end{figure*}
\subsection{Two detectors}

The no-go theorem described in \cite{Simidzija:2017kty,PhysRevD.97.125002} states that ``a pair of UDW detectors with Dirac-delta switching functions and arbitrary spatial profiles and coupling strengths cannot harvest entanglement 
from a coherent state of a massless scalar field.''  Therefore, the bandlimit---which is mathematically equivalent to a spatial profile---will not make a difference on the entanglement harvested (none).

One can still look at the transition probability of the second detector, detector $B$, which depends on the effective field commutator through the parameter $\tht$.  In figure \ref{fig:deltaThetaAndPB}(a), it is shown that $\tht$, when plotted as a function of the bandlimit exhibits damped oscillates with frequencies at $(S \pm T)$.  By studying the parameter space, we have found that when the spatial profile of the detector is made smaller, the oscillations decay more slowly, and in the limit of point-like detectors, $a\to0$, the oscillations do not decay.  In this limit, $\tht$ reduces to %
\eqn{
  \lim_{a\to0} \tht = \frac{\la^2\si^2\La}{2\pi S}\lb[\sinc\big(\La(T+S)\big)-\sinc\big(\La(T-S)\big)\rb]
  \label{eq:ato0theta}
}
which is simply the difference of the one-dimensional Fourier transforms of the rectangle function $\Pi(k/(2\Lambda))$ evaluated at $(T+S)$ and $(T-S)$.

The oscillations in $\theta$ are observable in the transition probability of the second detector when the bandlimit is small, and as shown in the inset of figure \ref{fig:deltaThetaAndPB}(b), these can result in a transition probability greater than~$1/2$, the largest transition probability of a single detector.  However, when the bandlimit is increased, the transition probability exponentially approaches its non-bandlimited quantity at the same rate as a single detector, and the oscillations resulting from $\theta$ are exponentially suppressed.  If the bandlimit is small, then a second detector with instantaneous switching can be used to give some insight into the field commutator between the locations of the two detectors, but if the bandlimit is large, then a second detector will not provide any advantage.

\section{Conclusion}
\label{sec:Conclusion}

We have studied the entanglement harvesting protocol and response of two UDW detectors interacting with a conventionally bandlimited $(3+1)$-dimensional scalar field, where modes with $|\bd{k}|\ge\La$ do not exist.  We find that the application of this cutoff is equivalent to modifying the spatial profile of the UDW detector by convolving it with the dimensionally appropriate Fourier transform of the rectangle function. This interpretation results in a non-local detector interacting with a non-bandlimed scalar field.  On the other hand, if the detectors were interpreted as local, then the degrees of freedom of the field are non-local. The two perspectives are entirely equivalent.

When a point-like detector couples to the bandlimited field with Gaussian switching, we find to lowest order in the coupling strength that the probability of excitation is reduced compared to the non-bandlimited value. The effect only becomes pronounced, however, when the bandlimit is very small.  We find, for all values of the bandlimit, when the detector is prepared in the excited state, the de-excitation probability increases linearly with increasing values of the energy gap of the detector, until the energy gap is slightly smaller than the bandlimit of the field.  For larger energy gaps, the de-excitation probability exponentially falls off to nearly zero. The interpretation of this is that, to lowest order, the detector must emit a single phonon in order to decay %
, but if the energy gap is larger than the bandlimit, the emitted photon cannot propagate in the field.

We also note that it is possible for a pair of point-like UDW detectors, coupling with Gaussian switching, to harvest entanglement from the vacuum of the bandlimited scalar field at larger separation distances than in the ordinary case of a field without a bandlimit.
The enhancement results from oscillations in the imaginary part of the coherence term of the two detectors, which can be interpreted as resulting from the overlap of the non-local effective spatial profile of the detectors.  By taking advantage of the enhancement in entanglement harvesting, one can in principle take an array of pairs of detectors, with specific energy gaps and separations, and put bounds on the bandlimit of a quantum field.

Finally, we use a Dirac-delta switching function to couple two detectors with Gaussian spatial profiles to the field at a single instance in time, which allows for non-perturbative solutions.  Again, we find that the transition probability of the first detector is lower when the field is bandlimited, and the effect is much more apparent when the bandlimit is small.  Surprisingly, the transition probability of a detector is most sensitive to the bandlimit when the detector's width is small---but not too small. We also find that the transition probability of the second detector has a similar dependence on the bandlimit: any changes are due to the effective field commutator between the two detectors.

Although it has a very small effect on the transition probability of the second detector, the value of the field commutator has significant dependence on the bandlimit, even when the latter is large. We expect that other setups that are highly dependent on the field commutator between two detectors will also be highly sensitive to the value of the bandlimit.  One such application is ``quantum collect calling'' \cite{jonsson:2015prl} where it possible to signal using a massless scalar field such that no energy is transmitted from the sender to the receiver.  Since this scheme significantly depends on the commutator of the field between the sender and receiver, it would be interesting further work to apply quantum collect calling to bandlimited QFT.

This work may also have applications in superconducting circuits where it was shown \cite{McKay:2017pra} that, in the ultrastrong coupling regime with non-adiabatic switching, it is in principle possible to gain knowledge about the ultraviolet cutoff behaviour of the open transmission line by studying the probabilities of spontaneous vacuum excitation and spontaneous emission of a superconducting qubit.

\acknowledgments
We  %
thank
Paul Alsing%
, 
Valentina Baccetti%
, 
Achim Kempf%
, 
Robert Mann%
, 
Keith Ng%
, 
Denis Seletskiy%
, 
Nadine Stritzelberger%
,
Erickson Tjoa%
, 
Daniel Grimmer%
,
and 
Eduardo Mart\'{i}n-Mart\'{i}nez
for useful discussions and comments on various aspects of this work. L.J.H.\ acknowledges support from the Natural Sciences and Engineering Research Council of Canada.  This work was supported by the Australian Research Council Centre of Excellence for Quantum Computation and Communication Technology (Project No.\ CE170100012), by the Australian Research Council Discovery Program (Project No.\ DP200102152), and by the U.S. Air Force Research Laboratory Asian Office of Aerospace Research and Development (Grant No.\ FA2386-16-1-4020). 

\bibliographystyle{apsrev4-2_title}
 \bibliography{references2}

\end{document}